\documentclass[traditabstract]{aa_nolineno}
\bibpunct{(}{)}{;}{a}{}{,} 
\usepackage{graphicx}
\usepackage{mathrsfs}
\usepackage[varg]{txfonts}
\usepackage[T1]{fontenc}
\usepackage{color}
\usepackage{stfloats}

\begin{document}

\title{Size and kinematics of the \ion{C}{iv} broad emission line region from microlensing-induced line profile distortions in two gravitationally lensed quasars}
\author{Damien Hutsemékers\inst{1,}\thanks{Research Director F.R.S.-FNRS} 
   \and Dominique Sluse\inst{1}
   \and Đorđe Savić\inst{1}
       }
\institute{
    Institut d'Astrophysique et de G\'eophysique,
    Universit\'e de Li\`ege, All\'ee du 6 Ao\^ut 19c, B5c,
    4000 Li\`ege, Belgium
    }
%
%
\titlerunning{Size and kinematics of the quasar \ion{C}{iv} broad emission line region from microlensing} 
\authorrunning{D. Hutsem\'ekers et al.}
\abstract{
Microlensing of the broad emission line region (BLR) in gravitationally lensed quasars produces line profile distortions that can be used to probe the BLR size, geometry, and kinematics. Based on single-epoch spectroscopic data, we analyzed the \ion{C}{iv} line profile distortions due to microlensing in two quasars,  SDSS~J133907.13$+$131039.6 (J1339) and SDSS~J113803.73$+$031457.7 (J1138), complementing previous studies of microlensing in the quasars Q2237$+$0305 and J1004$+$4112. J1339 shows a strong, asymmetric line profile deformation, while J1138 shows a more modest, symmetric deformation, confirming the rich diversity of microlensing-induced spectral line deformations. To probe the \ion{C}{iv} BLR, we compared the observed line profile deformations to simulated ones. The simulations are based on three simple BLR models, a Keplerian disk (KD), an equatorial wind (EW), and a polar wind (PW), of various sizes, inclinations, and emissivities. These models were convolved with microlensing magnification maps specific to the microlensed quasar images, which produced a large number of distorted line profiles. The models that best reproduce the observed line profile deformations were then identified using a Bayesian probabilistic approach. We find that the line profile deformations can be reproduced with the simple BLR models under consideration, with no need for more complex geometries or kinematics. The models with disk geometries (KD and EW) are preferred, while the PW model is definitely less likely. In J1339, the EW model is favored, while the KD model is preferred in Q2237$+$0305, suggesting that various kinematical models can dominate the \ion{C}{iv} BLR. For J1339, we find the \ion{C}{iv} BLR half-light radii to be $r_{1/2} =$ 5.1 $^{+4.6}_{-2.9}$  light-days and $r_{1/2} =$ 6.7 $^{+6.0}_{-3.8}$ light-days from spectra obtained in 2014 and 2017, respectively. They do agree within uncertainties. For J1138, the amplitude of microlensing is smaller and more dependent on the macro-magnification factor. From spectra obtained in 2005 (single epoch), we find $r_{1/2} =$ 4.9 $^{+4.9}_{-2.7}$  light-days and $r_{1/2}= $ 12 $^{+13}_{-8}$ light-days for two extreme values of the macro-magnification factor. Combining these new measurements with those previously obtained for the quasars Q2237$+$0305 and J1004$+$4112, we show that the BLR radii estimated from microlensing do follow the \ion{C}{iv} radius--luminosity relation obtained from reverberation mapping, although the microlensing radii seem to be systematically smaller, which could indicate either a selection bias or a real offset.
}
\maketitle
%
%
%

\section{Introduction}
\label{sec:intro} 

The properties of the broad emission line region (BLR) that characterizes active galactic nuclei (AGNs) are most often investigated using reverberation mapping, a technique that measures the time-lag response of the broad emission lines (BELs) after a variation in the ionizing continuum \citep{1982Blandford, 1993Peterson, 2021Cackett}.  This time lag is directly related to the size of the BLR. In the best cases, it can be measured as a function of the velocity across the emission line profiles, providing information on the geometry and kinematics of the BLR. The geometry is generally a thick disk viewed with a small to moderate inclination with respect to the line of sight (10$-$40\degr); it can show various kinematical signatures, mainly virialized motions, inflows or outflows \citep[e.g.,][]{2009Bentz, 2010Bentz, 2014Pancoast, 2016Dua, 2017Griera, 2018Xiao, 2018Williams, 2019Zhang, 2020Williams, 2021Bentz}. Assuming virial motion, the BLR radius, together with the gas velocity measured from the BEL Doppler width, gives the mass of the AGN supermassive black hole, a key parameter for understanding black hole growth and coevolution with host galaxies through cosmic time. Reverberation mapping also unveiled BLR radius--AGN luminosity ($R$-$L$) relations, which were expected from photoionization models \citep{2005Kaspi, 2013Bentz, 2018Lira, 2019Du, 2020Fonseca, 2021Kaspi, 2023Yu, 2023Shen}. Reverberation mapping studies have mainly focused on low-redshift AGNs and the H$\beta$ BEL. $R$-$L$ relations based on the \ion{Mg}{ii} and \ion{C}{iv} BELs observed in high-redshift luminous AGNs (quasars) are more difficult to accurately constrain. Indeed, due to the large BLR sizes and the time dilation associated with cosmological redshifts, reverberation mapping of quasars requires years to decades of monitoring. Independently, thanks to near-infrared interferometry, the P$\alpha$ and Br$\gamma$ BLRs have been spatially resolved in a handful of AGNs and found to be compatible with thick, rotating disks of clouds, in good agreement with the results from reverberation mapping \citep{2018Gravity, 2021Gravity, 2024Gravity}.

Microlensing of the BLR in gravitationally lensed quasars can provide independent measurements of the BLR size and kinematics, since the magnification of a source in the quasar core depends on its size: the smaller the source, the stronger the magnification (for a recent review, see \citealt{2024Vernardos}). While this technique can only be applied to lensed quasars, it can be used with single-epoch spectra of the different images, thus avoiding the years of monitoring needed by reverberation mapping to properly characterize the BLR of high-redshift objects. Several studies have shown that line profile distortions are commonly observed in the spectra of one or more images of lensed quasars; they are detected via comparisons to undistorted line profiles observed in at least one other image \citep{2004Richards, 2007Sluse, 2011ODowd, 2011Sluse, 2012Sluse, 2013Guerras, 2014Braibant, 2016Braibant, 2016Goicoechea, 2017Motta, 2018Fiana, 2020Popovic, 2021Fian}. When the time delay between the spectra coming from the different images of a lensed quasar is shorter than 40-50 days, the line profile deformations observed in one image can be attributed to microlensing rather than to intrinsic variations \citep{2012Sluse}. These line profile deformations are most often detected as red--blue or wings--core distortions, and explained in terms of differential magnification of spatially and kinematically separated subregions of the BLR. Line profile deformations have been predicted and computed in the framework of various models \citep{1988Nemiroff, 1990Schneider, 2001Popovic, 2002Abajas, 2007Abajas, 2004Lewis, 2011ODowd, 2011Garsden, 2014Simic, 2017Braibant}, which can be used to infer the BLR geometry and kinematics.

Microlensing of the BELs in the lensed quasar Q2237$+$0305 provided an estimate of the BLR size that was in good agreement with the $R$-$L$ relations obtained from reverberation mapping \citep{2005Wayth, 2011Sluse}. By simulating the effect of microlensing on simple models of the BLR, \cite{2011ODowd} found that the observed microlensing signature favors a gravitationally dominated BLR kinematics. Using single-epoch spectra of a sample of lensed quasars grouped into low- and high-luminosity subsamples, and assuming a Gaussian luminosity profile for the BLR, \citet{2013Guerras} estimated the size of the low- and high-ionization BLRs. They found that the high-ionization BLR is smaller than the low-ionization BLR, with a radius--luminosity dependence in agreement with $R$-$L$ relations from reverberation mapping. These results were refined and confirmed by \citet{2018Fiana,2021Fian} based on multi-epoch observations.

To further probe the geometry and kinematics of the BLR in individual objects based on single-epoch spectroscopic data,  we computed the effect of gravitational microlensing on the BEL profiles and the underlying continuum, using representative BLR models and microlensing magnification maps specific to the lensed quasars \citep{2017Braibant}. We then developed a Bayesian probabilistic approach to select the models that best reproduce the observed line profile deformations \citep{2019Hutsemekers}. We find that microlensing of a BLR with a disk geometry best reproduces the distortions of the H$\alpha$ and \ion{C}{iv} lines observed in the lensed quasars HE0435$-$1223, Q2237$+$0305, and J1004$+$4112 \citep{2019Hutsemekers, 2021Hutsemekers, 2023Hutsemekers, 2024Savic}. The size of the BLR was estimated and found to be either in good agreement with, for Q2237$+$0305, or smaller than, for J1004$+$4112, the values expected from the reverberation mapping $R$-$L$ relations. Measurements of the BLR size in the quasars Q0957$+$561 and J1004$+$4112 were recently obtained by \cite{2023Fian,2024Fian} based on a different method; for J1004$+$4112, they are in good agreement with our estimate.

In this paper we investigate, using single-epoch spectroscopic data, the size and kinematics of the \ion{C}{iv} highly ionized BLR in two gravitationally lensed quasars for which clear emission line profile differences are observed between some images, indicating microlensing of the BLR. The inferred BLR properties are compared with those of the previously studied objects, Q2237$+$0305 and J1004$+$4112, in particular in the context of the $R$-$L$ relations.

\section{Targets and data}
\label{sec:data}

The quasars SDSS~J133907.13$+$131039.6 (hereafter J1339) and SDSS~J113803.73$+$031457.7 (hereafter J1138) are the two lensed quasars under investigation.

J1339 shows two images A and B separated by 1\farcs7  \citep{2009Inada}.  The source is at redshift $z_s$ = 2.231 and the lens is an early type galaxy at redshift $z_l$ = 0.609 \citep{2014Shalyapin}. Image B is strongly affected by microlensing, with clear line profile distortions relative to image~A \citep{2016Goicoechea,2021Shalyapin}. The distortions are persistent over a period of at least a few years, much longer than the time delay of 47 days between the two images \citep{2021Shalyapin}, supporting the BLR microlensing interpretation against intrinsic variations observed with a delay between the two images. In the following, we consider the spectra of J1339 A\&B obtained on May 20, 2014, with the Gran Telescopio Canarias (GTC) equipped with OSIRIS, and on April 6, 2017, with the Very Large Telescope (VLT) equipped with X-shooter. These spectra are described in \cite{2021Shalyapin} and publicly available from the GLENDAMA archive\footnote{\tt https://grupos.unican.es/glendama/database/} \citep{2018GilMerino}. They cover the \ion{C}{iv} $\lambda$1549~\AA\ emission line with a signal-to-noise ratio S/N $\gtrsim 50$ that allows us to accurately measure the line profile distortions. A photometric monitoring revealed that, between 2014 and 2017, J1339 exhibited an intrinsic brightness increase of 0.4 magnitude in the $r$-SDSS band \citep{2021Shalyapin}.

J1138 is a quadruply imaged quasar with a maximum image separation of 1\farcs46. It has a redshift $z_s$ = 2.438, and is lensed by a galaxy at $z_l$ = 0.445 \citep{2006Eigenbrod}. A spectrum obtained through images B-C shows a microlensing magnification of the \ion{C}{iv} line wings \citep{2012Sluse}. The predicted time delay in this system is on the order of 5 days \citep{2012Sluse}, supporting the interpretation of the spectral difference in terms of microlensing. The analysis of X-ray images obtained in 2007 \citep{2015Jimenezb} indicates that images A and B are microlensed, and not images C and D. In the following, we use the spectra of J1138 B\&C obtained on May 10, 2005, with the VLT equipped with the FORS1 instrument, and described in \cite{2012Sluse}. These spectra were also secured with S/N $\gtrsim 50$, so that the line profile distortions in the \ion{C}{iv} $\lambda$1549~\AA\ emission line can be precisely characterized.

\section{Broad emission line microlensing}
\label{sec:bels}

\begin{figure*}[t]
\resizebox{\hsize}{!}{\includegraphics*{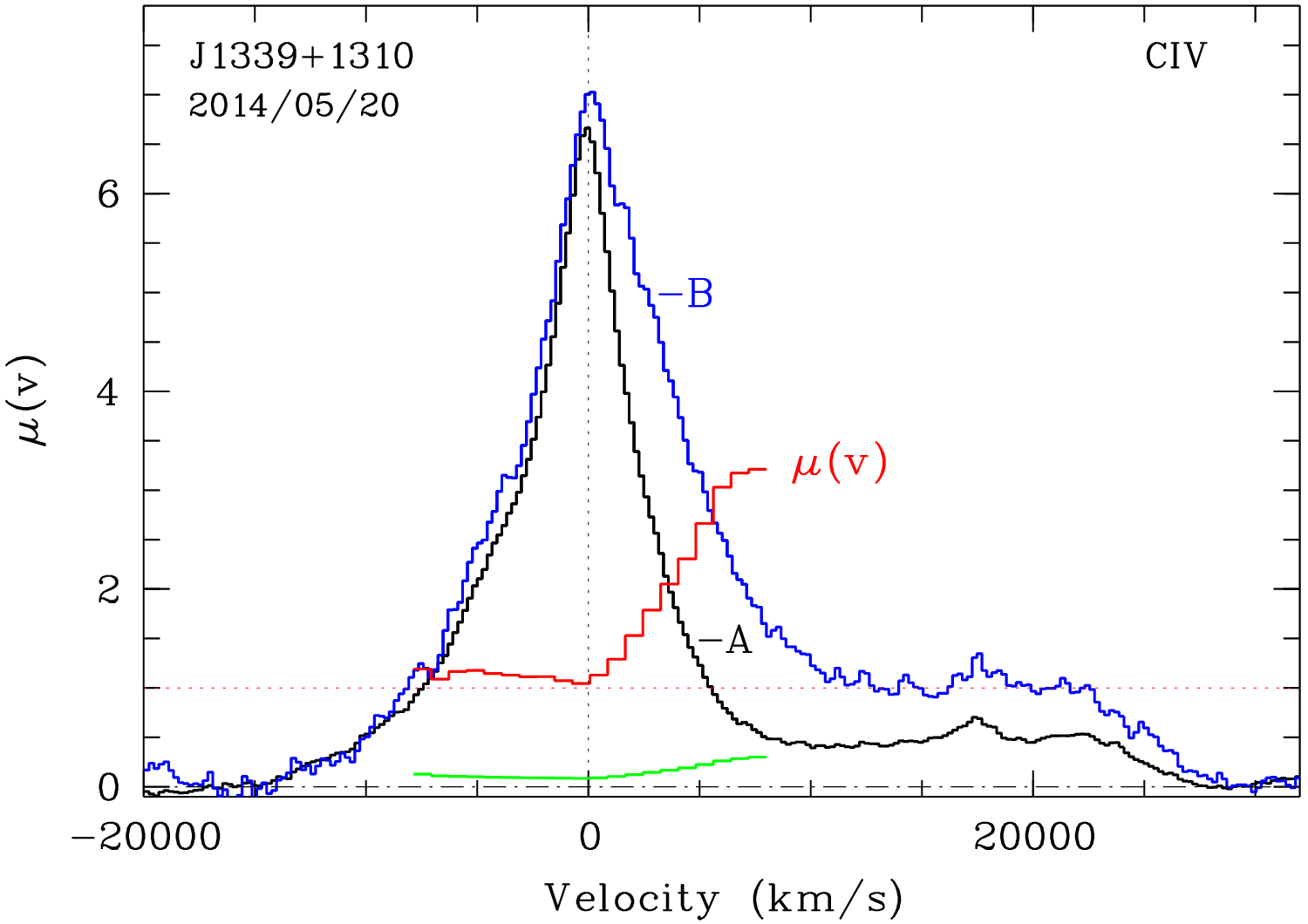}\includegraphics*{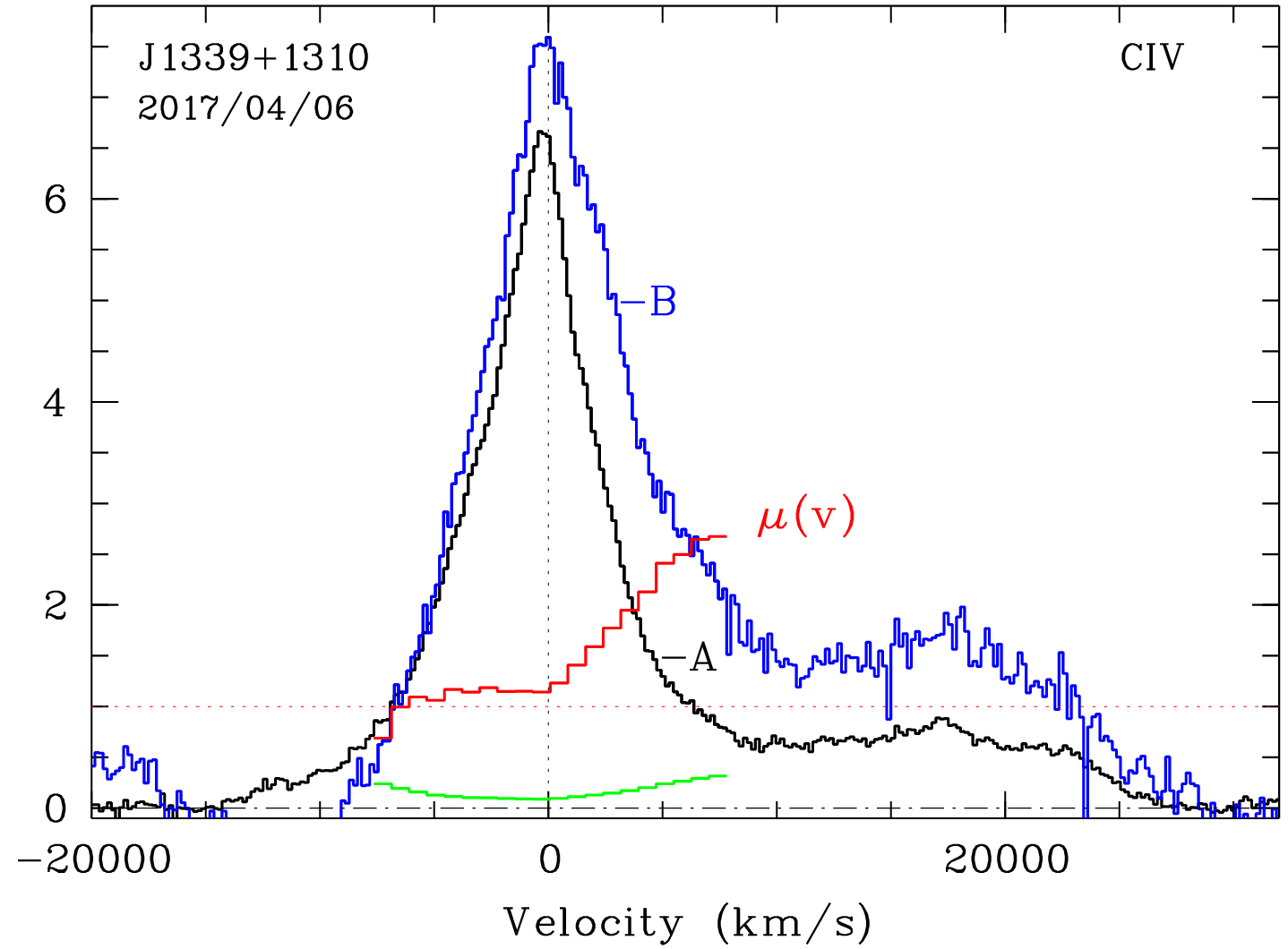}}\\
\caption{ $\mu(v)$ magnification profiles of \ion{C}{iv} (in red) computed from spectra of images B and A of J1339, simultaneously recorded.  Profiles computed from the spectra obtained in 2014 and 2017 are illustrated in the left and right panels, respectively. The $\mu(v)$ profiles are binned into 20 spectral elements, with the uncertainties shown in green. The superimposed line profiles from image B (in blue) and A (in black) are continuum-subtracted, corrected by the  $M(\text{\ion{C}{iv}})$ factor, and arbitrarily rescaled. The zero-velocity corresponds to the \ion{C}{iv} $\lambda$1549 wavelength at the source redshift.}
\label{fig:muv1}
\end{figure*}

\begin{figure*}[t]
\resizebox{\hsize}{!}{\includegraphics*{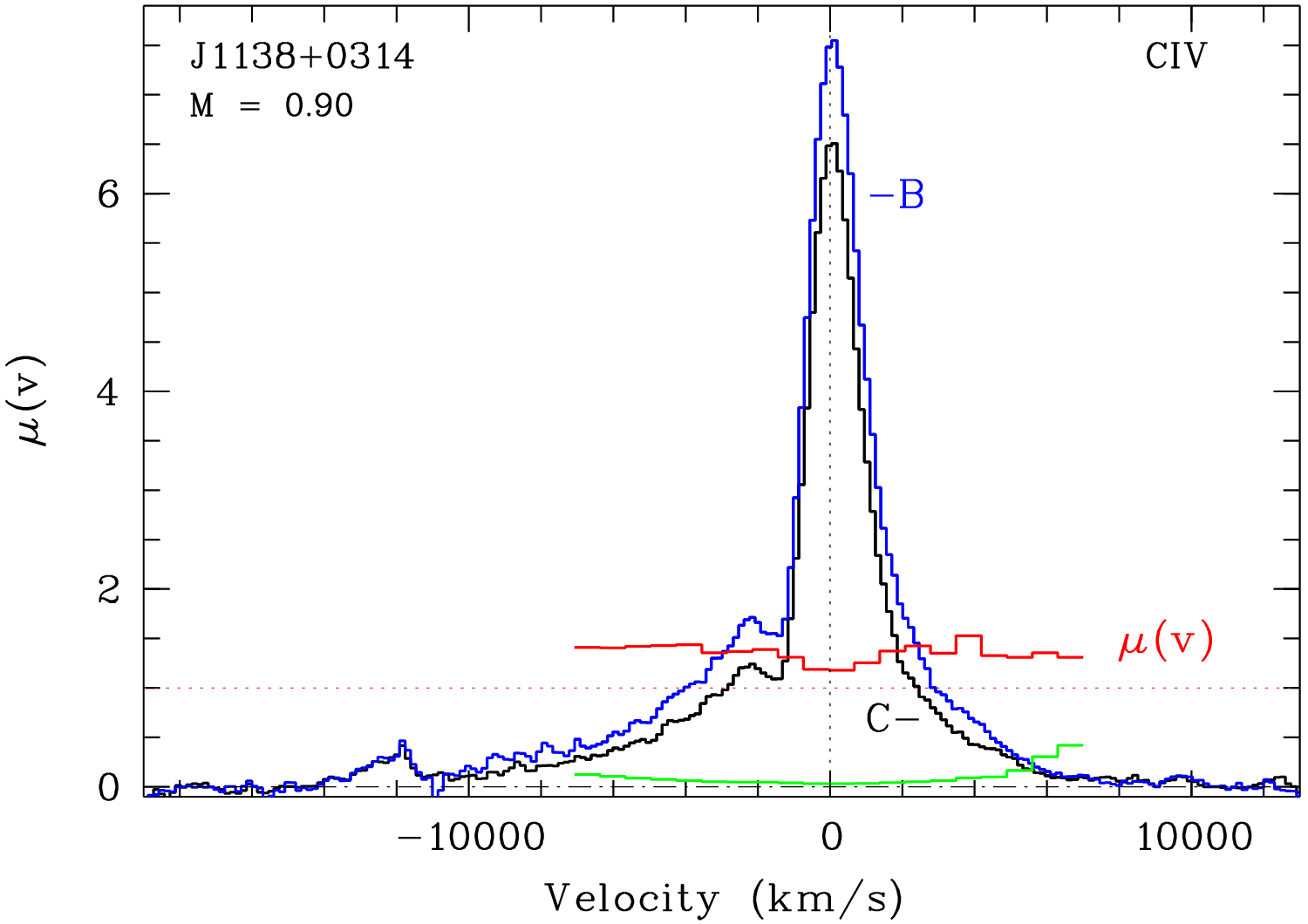}\includegraphics*{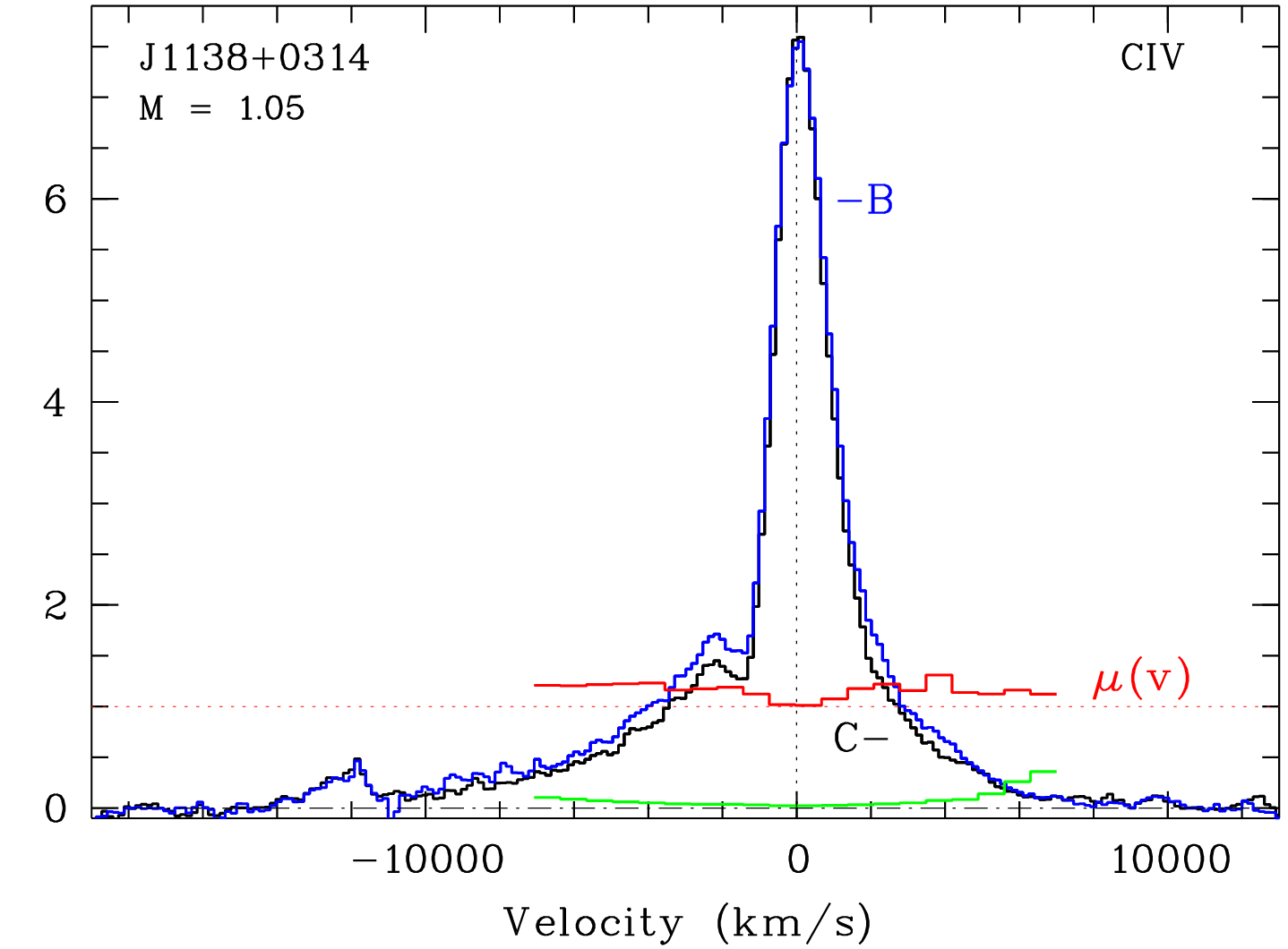}}\\
\caption{ $\mu(v)$ magnification profiles of \ion{C}{iv} (in red) computed from spectra of images B and C of J1138,  simultaneously recorded.  The $\mu(v)$ profiles are binned into 20 spectral elements, with the uncertainties shown in green.  The superimposed line profiles from image B (in blue) and C (in black) are continuum-subtracted, corrected by the $M(\text{\ion{C}{iv}})$ factor, and arbitrarily rescaled. Two values of the factor $M(\text{\ion{C}{iv}})$ are used:  $M(\text{\ion{C}{iv}})$ = 0.90 (left panel) and 1.05 (right panel).  The zero-velocity corresponds to the \ion{C}{iv} $\lambda$1549 wavelength at the source redshift.}
\label{fig:muv2}
\end{figure*}

\begin{table*}[]
\caption{Measured magnification and distortion indices.}
\label{tab:indices}
\renewcommand{\arraystretch}{1.2}
\centering
\begin{tabular}{lcccc}
\hline\hline
 Object - Date  & $\mu^{cont}$ & $\mu^{BLR}$ & WCI & RBI \\
\hline
J1339 - 2014  & 4.36$\pm$0.66 & 1.36$\pm$0.20 & 1.58$\pm$0.04 & 0.28$\pm$0.01 \\
J1339 - 2017  & 5.68$\pm$0.87 & 1.38$\pm$0.21 & 1.35$\pm$0.08 & 0.29$\pm$0.02 \\
J1138 - 2005  & 1.42$\pm$0.08 & 1.26$\pm$0.06 & 1.17$\pm$0.03 & 0.00$\pm$0.02 \\
\hline
\end{tabular}
\end{table*}

To characterize and quantify the line profile deformations induced by the microlensing effect, we considered the magnification profile $\mu(v)$, which is the macro-magnification-corrected ratio of the continuum-subtracted emission line flux densities observed in two different images. Denoting the emission line flux densities  $F^l_{\text{1}}$ and $F^l_{\text{2}}$, for images 1 (microlensed) and 2 (not microlensed), respectively, we write\begin{equation}
\mu \, (v) =  \frac{1}{M}  \frac{F^l_{\text{1}} \, (v) }{F^l_{\text{2}} \, (v)} \; ,
\label{eq:muv}
\end{equation}
where $M = M_{\text{1}} / M_{\text{2}}$ is the macro-magnification ratio of images 1~and~2, and $v$ the velocity computed from the \ion{C}{iv} line center at the source redshift. We also considered three indices integrated over the $F^l_{\text{1}} (v)$, $F^l_{\text{2}} (v)$, or  $\mu(v)$ profiles (see \citealt{2017Braibant} or \citealt{2021Hutsemekers} for exact definitions): (1)~$\mu^{BLR}$, which essentially quantifies the total magnification of the line, (2) the wings--core index (WCI); when different from one it indicates whether the whole emission line is, on average, more or less affected by microlensing than its center, and  (3), the red--blue index (RBI), which takes non-null values when the effect of microlensing on the blue and red parts of the line is asymmetric. A fourth index,  $\mu^{cont}$, measures the microlensing magnification of the continuum underlying the emission line.  When the spectra of both images are simultaneously recorded,  $\mu(v)$ and the indices are independent of quasar intrinsic variations that occur on timescales longer that the time delay between the two images. The measurement of the indices $\mu^{cont}$ and $\mu^{BLR}$, which characterize the strength of the micro-magnification, depend on the macro-magnification ratio $M$, while the RBI and WCI indices are independent of this ratio.

An accurate measurement of the macro-magnification ratio $M$ is thus needed to correctly estimate $\mu(v)$, $\mu^{cont}$, and $\mu^{BLR}$. Moreover, in Eq.~\ref{eq:muv}, $F^l_{\text{1}}$ and $F^l_{\text{2}}$ should be corrected for the differential extinction between images 1 and 2 that may arise from their different light paths through the lens galaxy. Since differential extinction equally affects the continuum and the lines, as opposed to microlensing, it can be conveniently incorporated into the factor $M$ instead of correcting $F^l_{\text{1}}$ and $F^l_{\text{2}}$. In this case, $M$ becomes wavelength dependent, $M (\lambda) =  (M_{\text{1}} / M_{\text{2}}) \times (\epsilon_{\text{1}} / \epsilon_{\text{2}})$, where $\epsilon_{\text{1}} (\lambda)$ and $\epsilon_{\text{2}} (\lambda)$ represent the transmission factors of the light from images 1 and 2, respectively. $M (\lambda)$ is then estimated at the wavelength of the line under consideration.

In the lensed system J1339, image B is microlensed, and \cite{2016Goicoechea} estimated $M_{\text{B}} / M_{\text{A}} = 0.175 \pm 0.015$,  $\epsilon_{\text{B}} = 1$, and $\epsilon_{\text{A}} = 0.56 \pm 0.09, 0.77 \pm 0.06$, and $0.86 \pm 0.04$ at the emission wavelengths $\lambda$ = 1350, 3000, and 5100~\AA , respectively \citep[see also][]{2021Shalyapin}. These values were obtained under the assumption that the line cores of some BELs originate in extended regions not affected by microlensing. In the 2017 spectrum of J1339, the [\ion{O}{iii}] $\lambda$~4959, 5007~\AA\ forbidden lines are detected. These lines come from the extended narrow line region so that their flux ratio can be used to compute $M_{\text{B}} / M_{\text{A}}$ without any assumption on the line cores. Taking into account the extinction in image A, we obtain $M_{\text{B}} / M_{\text{A}}$ = 0.20 $\pm$ 0.03 from the [\ion{O}{iii}] line ratio, in excellent agreement with the value derived by  \cite{2016Goicoechea}. Interpolating $\epsilon_{\text{A}}$ at the wavelength of \ion{C}{iv}, we derive $M(\text{\ion{C}{iv}}) = 0.33\pm0.05$, used in the computation of $\mu(v)$, $\mu^{cont}$, and $\mu^{BLR}$.

In the lens system J1138, \cite{2012Sluse} used the Macro-micro decomposition (MmD) method to derive $M$ at the \ion{C}{iv} wavelength. This method provides a correct measurement of $M$ under the assumption that the continuum and the emission line are either magnified or demagnified, and that at least a small part of the emission line is not microlensed \citep{2010Hutsemekers}. We emphasize that the MmD provides a direct measurement of $M(\text{\ion{C}{iv}})$, without separating the true macro-magnification ratio from differential extinction. With image B microlensed and image C not microlensed, $M(\text{\ion{C}{iv}}) = 0.90\pm0.05$ \citep{2012Sluse}. The fact that this value is significantly different from $M$ measured in the H and Ks band, $M(\text{H})$ = 0.84$\pm$0.01 and $M(\text{Ks})$ = 0.73$\pm$ 0.07 \citep{2012Sluse}, suggests that differential extinction is also at work in this system, providing a simple explanation to the difference of $M$  measured in the H and Ks bands. Alternatively, we could assume  $M(\text{\ion{C}{iv}})$ =  $M(\text{Ks})$ and no differential extinction, but this would imply a strong magnification of the \ion{C}{iv} emission line core, which is not very plausible. Moreover, the difference of $M$ measured in the H and Ks bands would not be easily explained. The value $M(\text{\ion{C}{iv}}) = 0.90\pm0.05$ was obtained from the simultaneous disentangling of the \ion{C}{iv} and \ion{C}{iii}] lines. Relaxing this constraint, values of $M(\text{\ion{C}{iv}})$ up to $1.05\pm0.05$ are still acceptable. Given these uncertainties, we considered both values in subsequent analyses, that is, $M(\text{\ion{C}{iv}}) = 0.90\pm0.05$ and $M(\text{\ion{C}{iv}}) = 1.05\pm0.05$ in the computation of $\mu(v)$, $\mu^{cont}$, and $\mu^{BLR}$.

Figures~\ref{fig:muv1} and~\ref{fig:muv2} show the $\mu(v)$ magnification profiles of the \ion{C}{iv} emission line for the quasars J1339 and J1138, respectively. The $\mu(v)$ profile was computed according to Eq.~\ref{eq:muv}, with $M(\text{\ion{C}{iv}}) = 0.33\pm0.05$ for J1339, and $M(\text{\ion{C}{iv}}) = 0.90\pm0.05$ and $1.05\pm0.05$ for J1138. The line flux densities of images B and A of J1339 (B and C for J1138) are continuum-subtracted. For J1339, the continuum was measured in two windows on each side of the line profile, $[-18,-15]$ and $[+27.5,+30]$ 10$^3$~km~s$^{-1}$, and interpolated by a straight line under the line profile.  The flux drop in the velocity range $[-14,-9]$ 10$^3$~km~s$^{-1}$ seen in the 2017 spectrum of image B is most likely an artifact and thus not taken into account in the continuum subtraction. For J1138, the two windows are $[-18,-16]$ and $[+10,+12]$ 10$^3$~km~s$^{-1}$. As a flux ratio, $\mu(v)$ can be extremely noisy in the wings of the emission lines where the flux density reaches zero, so that it is necessary to cut the faintest parts of the line wings.  We thus only considered the parts of the line profiles whose flux density is above $l_{\rm cut} \times F_{\rm peak}$, where $F_{\rm peak}$ is the maximum flux in the line profile and $l_{\rm cut}$ is fixed to 0.1 for J1339, and to 0.03 for J1138. This cutoff also allows us to discard artifacts that affect the far wings. To increase the signal-to-noise ratio, we binned $\mu(v)$ into 20 spectral elements, which also corresponds to the spectral resolution of the line profiles produced by the microlensing simulations (Sect.~\ref{sec:models}). The four indices that summarize the microlensing effect on the \ion{C}{iv} emission line are reported in Table~\ref{tab:indices}. For J1138, $\mu^{cont}$ and $\mu^{BLR}$ are given for $M(\text{\ion{C}{iv}}) = 0.90\pm0.05$. With $M(\text{\ion{C}{iv}}) = 1.05\pm0.05$, these values should be divided by a factor 1.05/0.90 = 1.17.

As seen in Fig.~\ref{fig:muv1} and Table~\ref{tab:indices}, the microlensing effect in J1339 is very strong, in both the continuum and the line. It is also strongly asymmetric, showing a magnification of the red part of the line profile that increases with the velocity, and no significant effect on the blue part. The microlensing effect in J1138 is very different: it is more modest, symmetric and it affects mostly the line wings with an essentially flat $\mu(v)$ profile, while the line core is less or not microlensed. These magnification profiles are also different from those observed in Q2237$+$0305 and J1004$+$4112 \citep{2021Hutsemekers, 2023Hutsemekers}, showing a nice diversity among the line profile distortions that can be induced by microlensing. 

\begin{figure*}[t]
\resizebox{\hsize}{!}{\includegraphics*{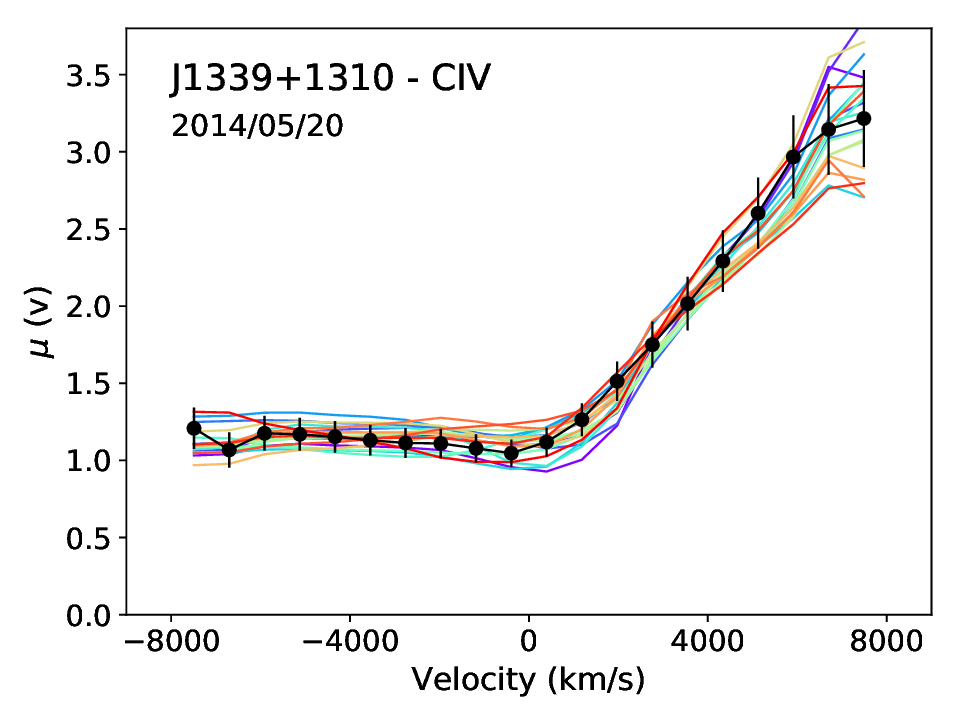}\includegraphics*{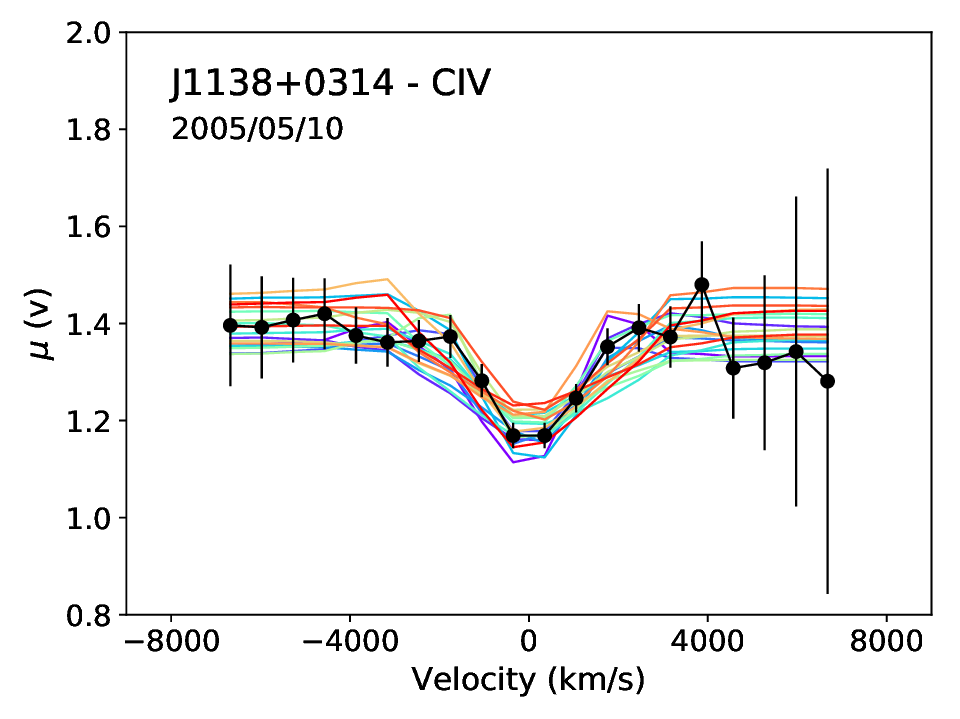}}\\
\caption{Examples of 20 simulated $\mu(v)$ profiles (in color) that fit the $\mu(v)$ magnification profiles (in black) of the \ion{C}{iv} emission line observed in J1339 (left) and J1138 (right). For J1339, the model is EW with $r_{\rm in} =  0.1 \, r_{\rm E} $, and  the magnification map oriented at $\theta$ = 60\degr . For J1138, the model is KD with $r_{\rm in} =  0.15 \, r_{\rm E} $, $\theta$ = 0\degr , and  $\mu(v)$ computed with $M(\text{\ion{C}{iv}})$ = 0.9. For both models, $i = 34\degr$, $q = 1.5$, and $r_s = 0.1 \, r_{\rm E}$.}
\label{fig:fitmuv}
\end{figure*}

\section{Microlensing simulations}
\label{sec:models}

We computed the effect of gravitational microlensing on the BEL profiles by convolving, in the source plane, the emission from representative BLR models with microlensing magnification maps. The microlensing simulations and the comparison to observations were carried out as described in \cite{2023Hutsemekers}. The method is essentially based on \cite{2017Braibant} where the models are detailed, and \cite{2019Hutsemekers}, where the probabilistic analysis is developed. In the following we briefly summarize the method. For details, we refer to the above cited papers.

For the BLR models, we considered a rotating Keplerian disk (KD), a biconical, radially accelerated polar wind (PW), and a radially accelerated equatorial wind (EW),  with inclinations with respect to the line of sight of $i$ = 22\degr, 34\degr, 44\degr, and 62\degr. Using the radiative transfer code STOKES \citep{2007Goosmann,2012Marin,2014Goosmann}, we produced 20 BLR monochromatic images which correspond to 20 spectral bins in the line profile. The BLR models were assumed to have an emissivity $\varepsilon = \varepsilon_0 \, (r_{\text{in}}/r)^q$ that either decreases sharply with increasing radius, $q=3$, or more slowly, $q=1.5$. We then attributed a range of sizes to these BLR models expressed in terms of the microlensing Einstein radius in the source plane, $r_E$.  For a lens of mass $\mathcal{M}$,
\begin{equation}
  \label{eq:re}
  r_E = \sqrt{4 \; \frac{G\mathcal{M}}{c^2} \frac{D_S D_{LS}}{D_L}} \, ,
\end{equation}
where $D_S$, $D_L$, and $D_{LS}$ are the source, lens, and lens-source angular diameter distances, respectively. For J1339, we considered 12 values of the BLR inner radius: $r_{\text{in}}$ = 0.025, 0.05, 0.075, 0.1, 0.125, 0.15, 0.175, 0.2, 0.25, 0.35, 0.5, and 0.75 $r_E$. For J1138, we used 15 $r_{\text{in}}$ values to cover a slightly broader range, adding $r_{\text{in}}$ = 0.01, 1.0, and 1.5 $r_E$ to the previous list. In all cases, the outer radius of the BLR is fixed to $r_{\text{out}} = 10 \, r_{\text{in}}$.

The continuum source is modeled as a disk of constant surface brightness (uniform disk), seen under the same inclination as the BLR. Since the effect of microlensing on circular disk models is rather insensitive to the surface brightness profile, the half-light radius being the primary parameter that controls the amplitude of the magnification \citep{2005Mortonson,2007Congdon}, we thus only considered uniform disks with outer radii ranging from $r_s =$ 0.01 to 5~$r_E$.

Modeling the effect of microlensing on the BLR and the continuum source is achieved using specific magnification maps computed with the \texttt{microlens} ray-tracing code \citep{1999Wambsganss}. The values of total convergence $\kappa$ and shear $\gamma$ at the image positions are based on the macro-models presented in \citet{2021Shalyapin} for J1339, and in \citet{2012Sluse} for J1138. \citet{2021Shalyapin} built a sequence of ten macro-models where the lens is made of a baryonic component tracing light, a circular dark matter halo, and an external shear. Each of the ten models differs by its fraction of stellar mass, which ranges between 0.1 and~1. These models predict a value of the stellar fractional mass density $\kappa_{\star}/\kappa$ at the image position. We elected two models corresponding to low and high stellar mass fractions. The values of convergence and shear for image B of J1339 are, for the first model, ($\kappa$, $\gamma$, $\kappa_{\star}/\kappa$) = (0.86, 0.45, 0.11), and, for the second model, ($\kappa$, $\gamma$, $\kappa_{\star}/\kappa$) = (0.63, 0.90, 0.52). The macro-model of J1138 consists of a singular isothermal ellipsoid + external shear, which predicts ($\kappa$, $\gamma$) = (0.54, 0.66) at the position of image B. There is no prediction on the stellar mass fraction from this model. Hence, we assumed two values of $\kappa_{\star}/\kappa$: 0.07 and 0.2. These values match the average mass fraction in compact objects at the location of lensed images \citep{2009Mediavilla,2015Jimeneza}. While the maps associated with the macro-models yield substantially different magnification distributions \citep{2014Vernardos}, the BLR properties we infer depend marginally on the specific map choice  (Sec.~\ref{sec:results}) so that a broader choice of map properties is not necessary. The maps extend over a $200 \, r_E \times 200 \, r_E$ area of the source plane and are sampled by $20000 \times 20000$ pixels. To mitigate the impact of the orientation of the symmetry axis of the BLR models relative to the caustic network, the maps are rotated clockwise by $\theta$ = 15\degr, 30\degr, 45\degr, 60\degr, 75\degr\ and 90\degr\ with respect to the BLR model axis. $\theta$ = 0\degr\ corresponds to the caustic elongation and shear direction perpendicular to the BLR model axis. After rotation, only the central $10000 \times 10000$ pixel part of the map is used.

Distorted line profiles are obtained by convolving, for a given BLR size, the magnification maps with the monochromatic images of the BLR, which depend on the model, the inclination, and the emissivity. Simulated $\mu(v)$ profiles are then computed for each position of the BLR on the magnification maps, generating $\sim 10^8$ simulated profiles per map and BLR model. The continuum-emitting region is treated in a similar way. The process is repeated for each size, inclination, and emissivity of the BLR models, for each size of the continuum source, and for each map orientation. The likelihood that the simulations reproduce the observables, $\mu^{cont}$ and the 20 spectral elements of $\mu(v)$, is then computed for each set of parameters characterizing the simulations. 

\begin{figure*}[t]
  \resizebox{\hsize}{!}{\includegraphics*{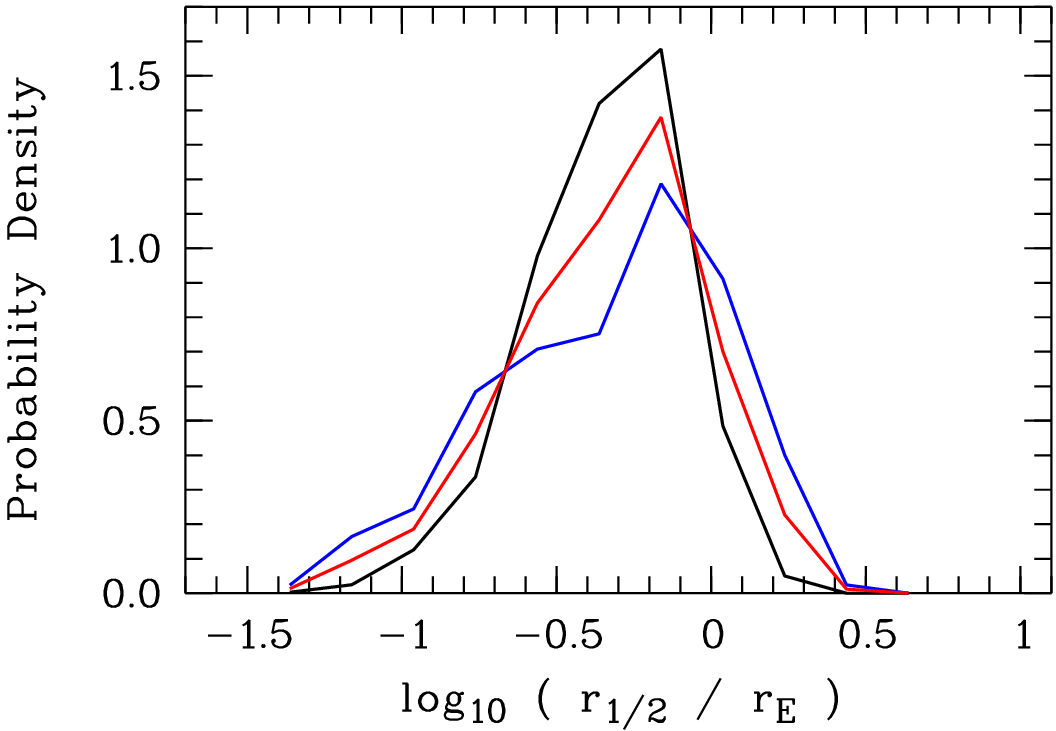}\includegraphics*{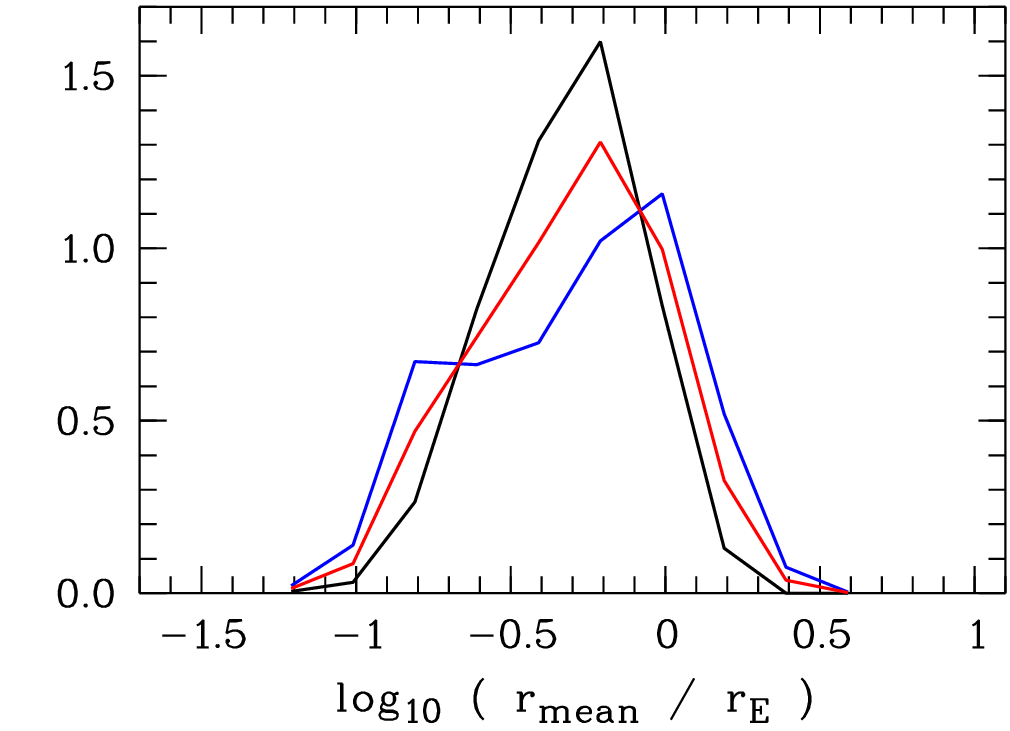}\includegraphics*{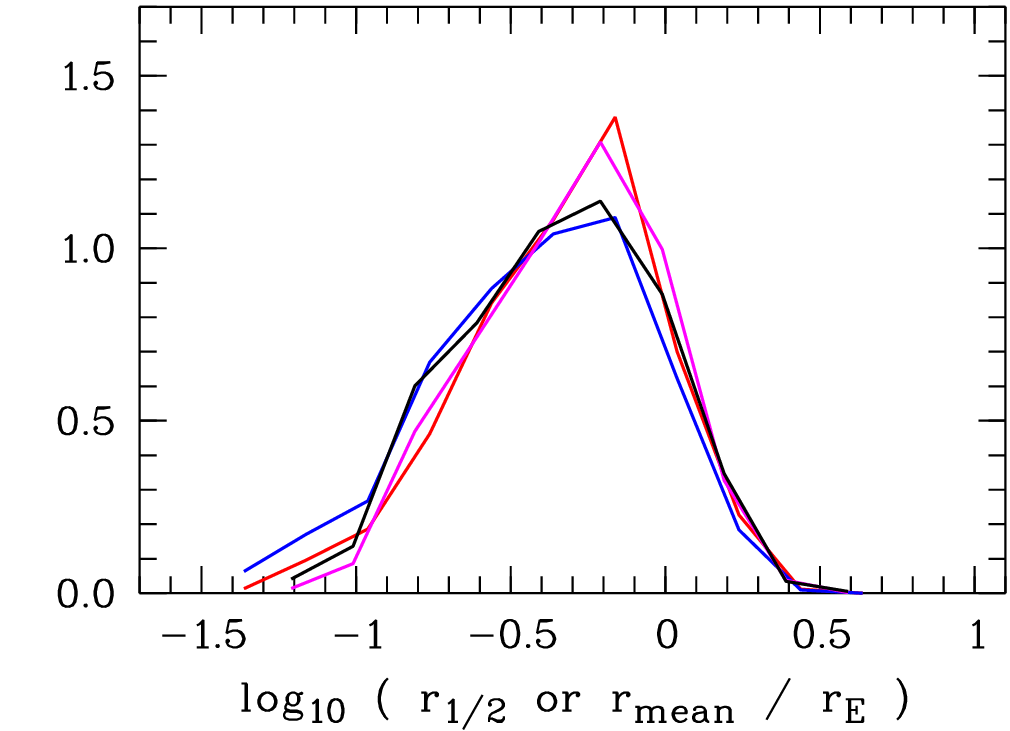}}\\
  \resizebox{\hsize}{!}{\includegraphics*{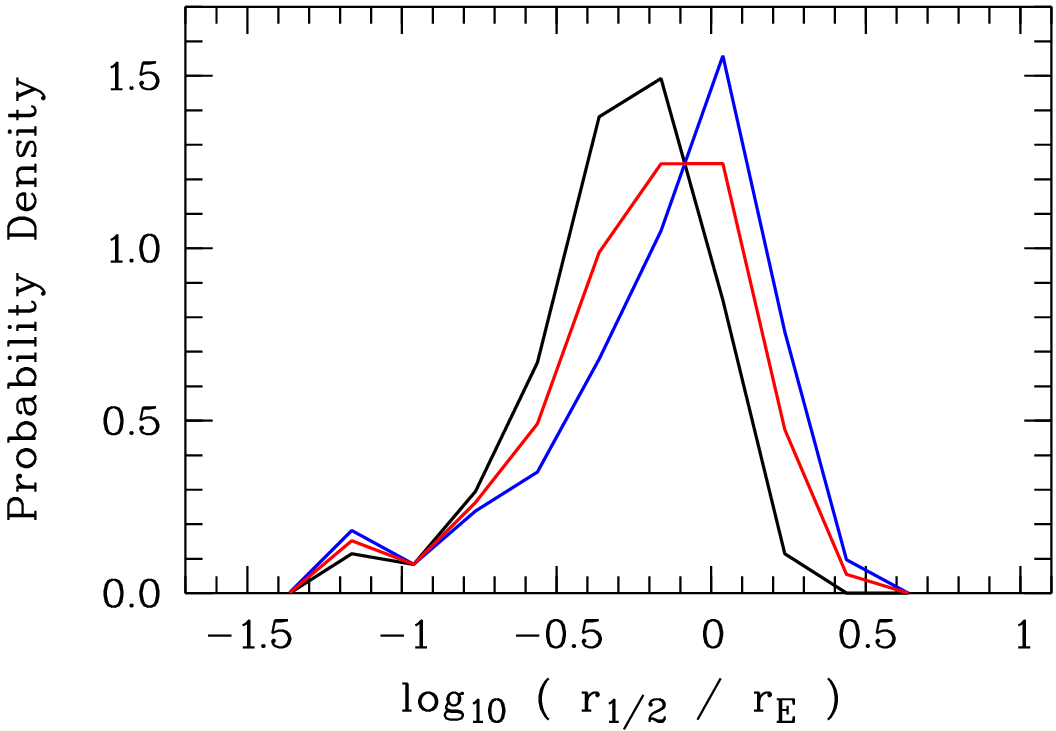}\includegraphics*{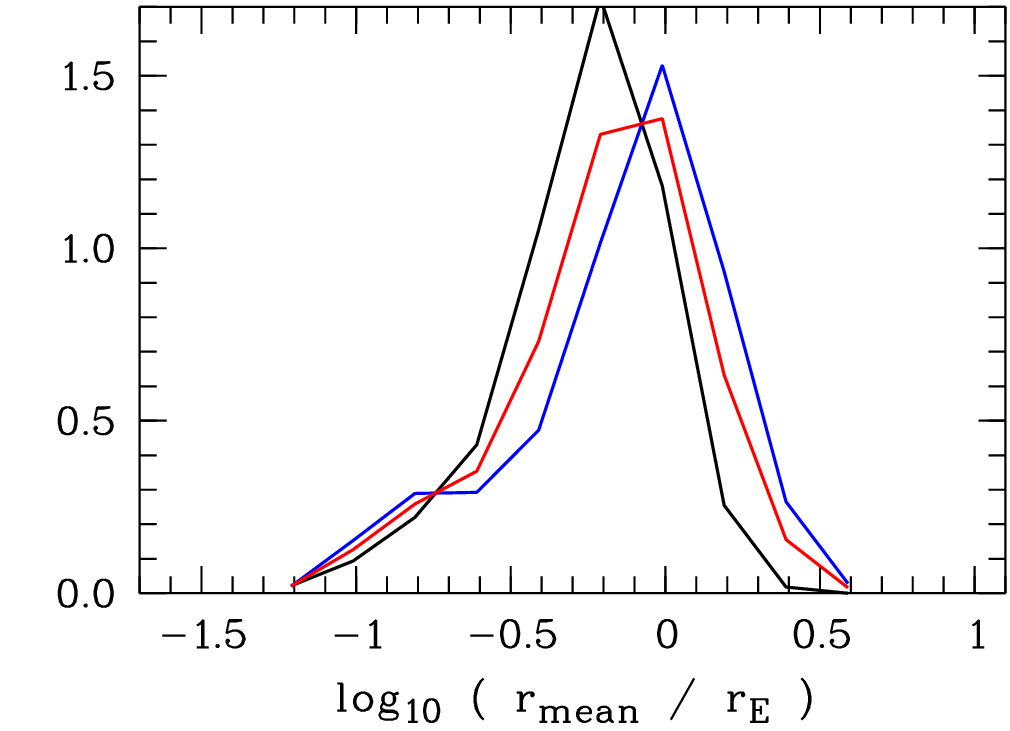}\includegraphics*{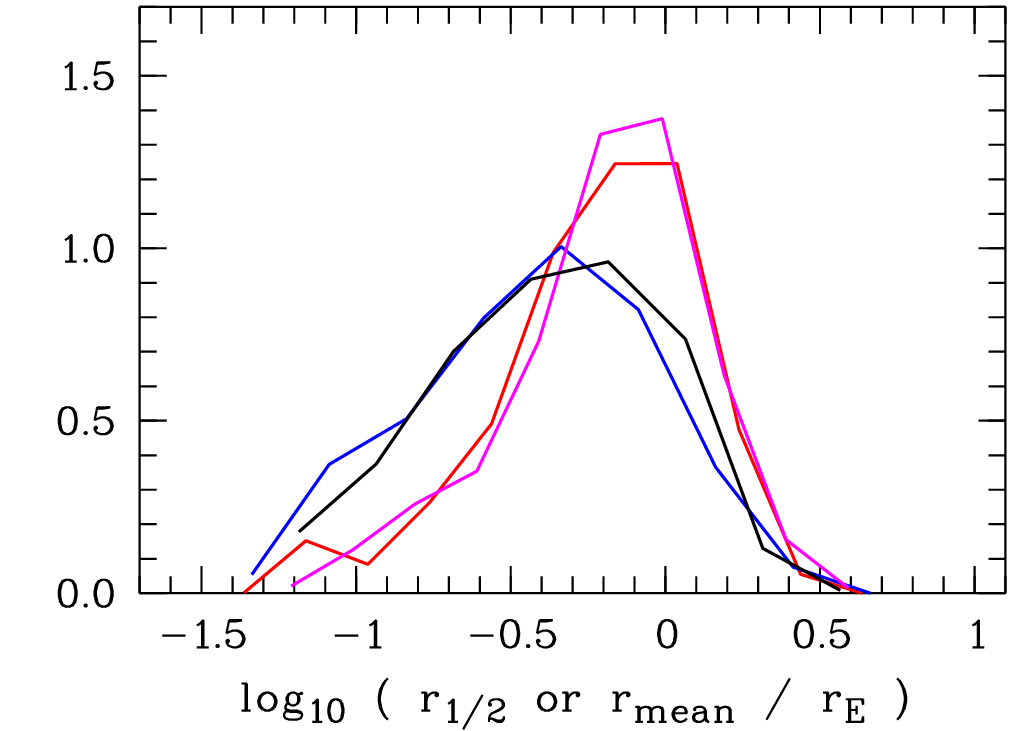}}
\caption{Posterior probability densities of the radius of the \ion{C}{iv} BLR in J1339. The BLR radius is expressed in Einstein radius units, with $r_E$ = 10.1  light-days for $\mathcal{M} =  0.3  \mathcal{M}_{\odot}$. The top panels show the 2014 data, and the bottom panels the 2017 data. {\it Left panels}: Probability densities of the half-light radius $r_{\text{1/2}}$ obtained with two magnification maps that were computed with different fractions of compact objects, $\kappa_{\star} / \kappa $ = 11\% (black) and $\kappa_{\star} / \kappa $ = 52\% (blue), and after marginalizing over the two maps (red).  {\it Middle panels}: Same as the left panels but for the flux-weighted mean radius $r_{\text{mean}}$. {\it Right panels}: Comparison of the probability densities, marginalized over the two maps, of the half-light radius (red and blue curves) and the flux-weighted mean radius (magenta and black curves), computed with constraints from the continuum source magnification (red and magenta curves) and without this constraint (blue and black curves).}
\label{fig:sizeblr1}
\end{figure*}
\begin{figure*}[t]
  \resizebox{\hsize}{!}{\includegraphics*{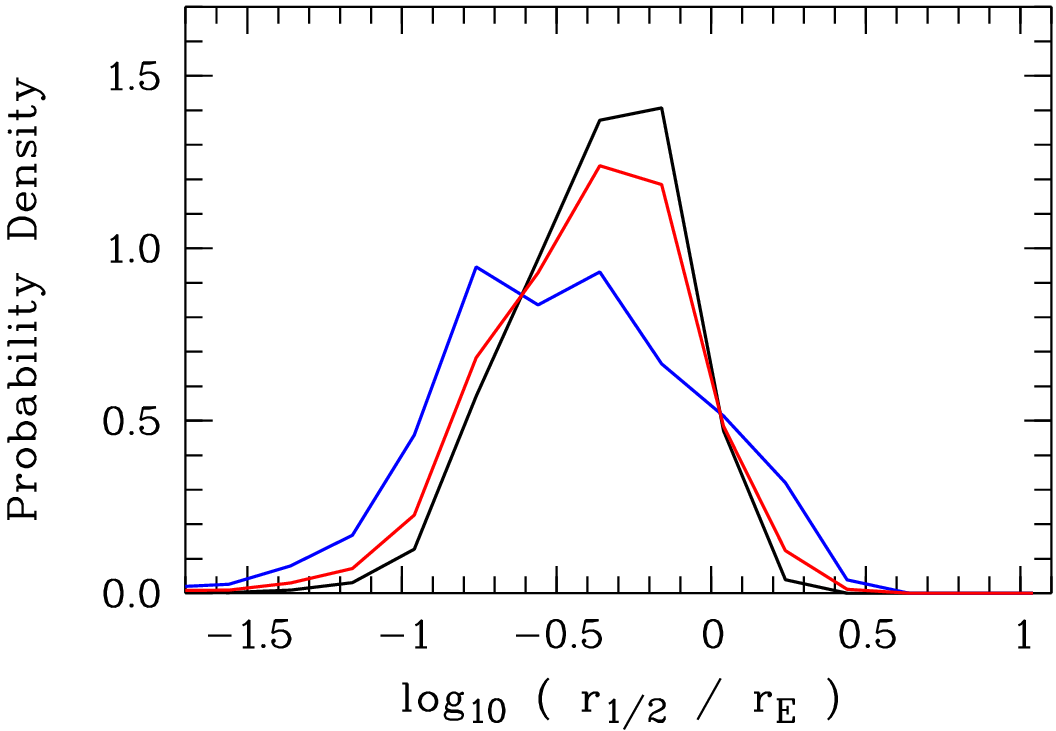}\includegraphics*{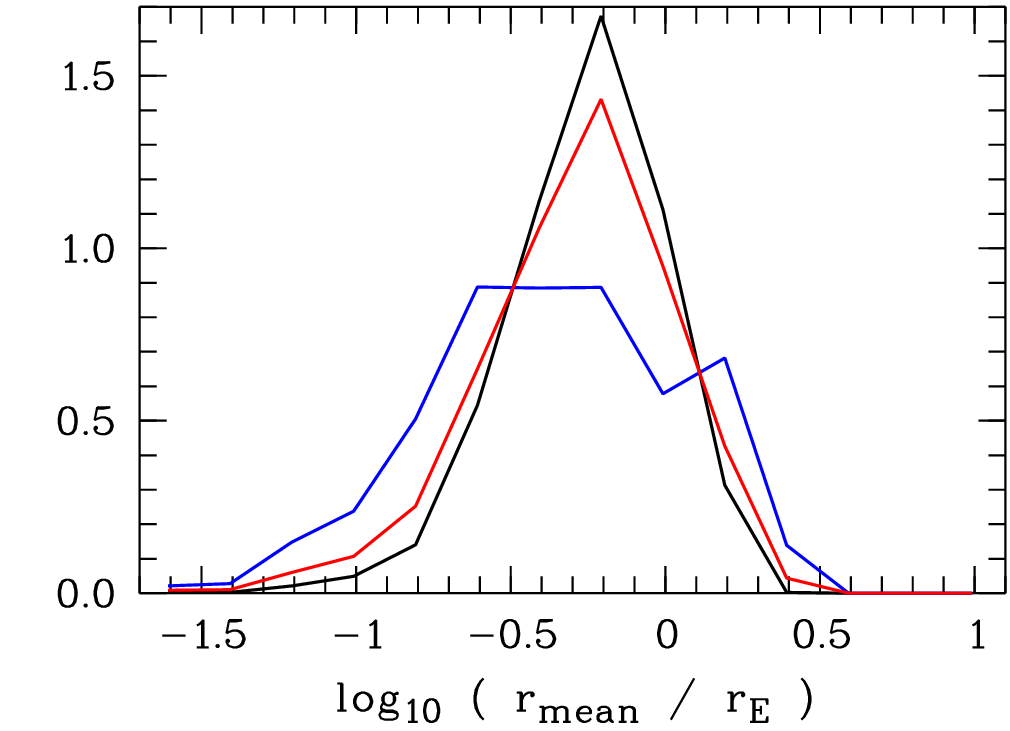}\includegraphics*{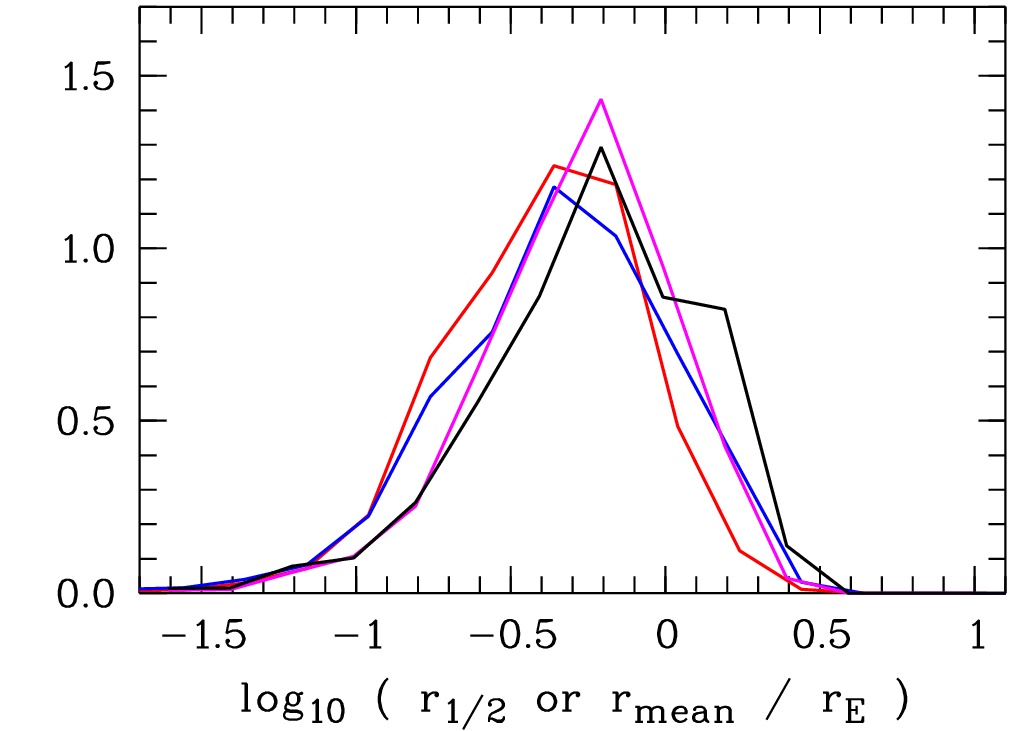}}\\
  \resizebox{\hsize}{!}{\includegraphics*{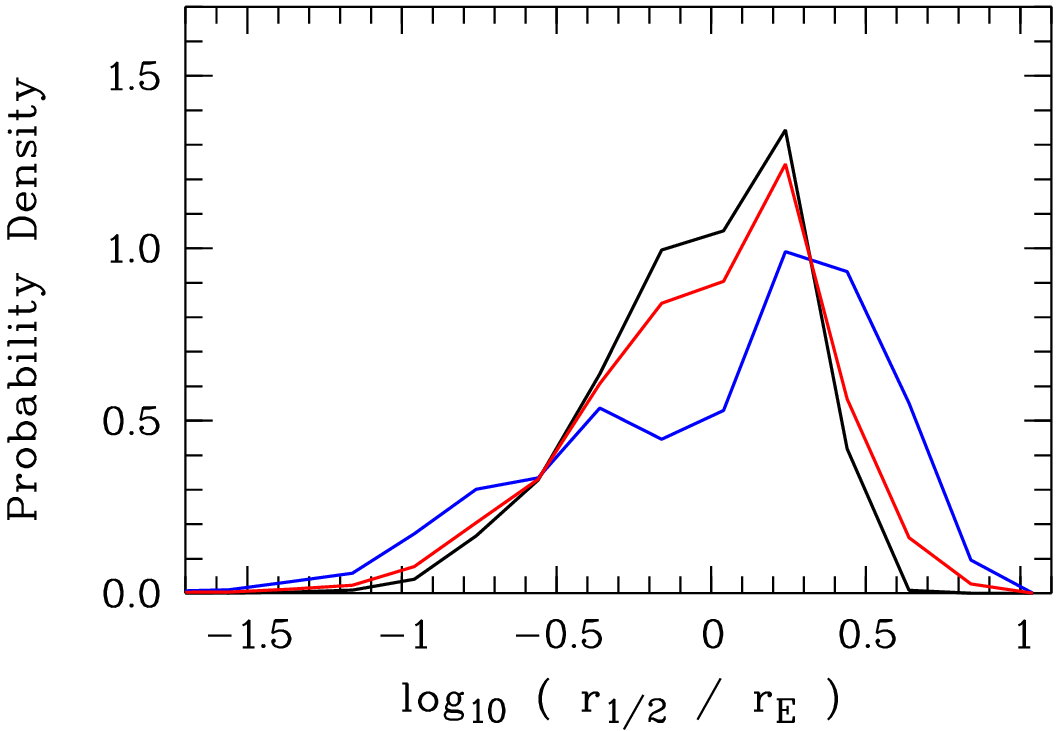}\includegraphics*{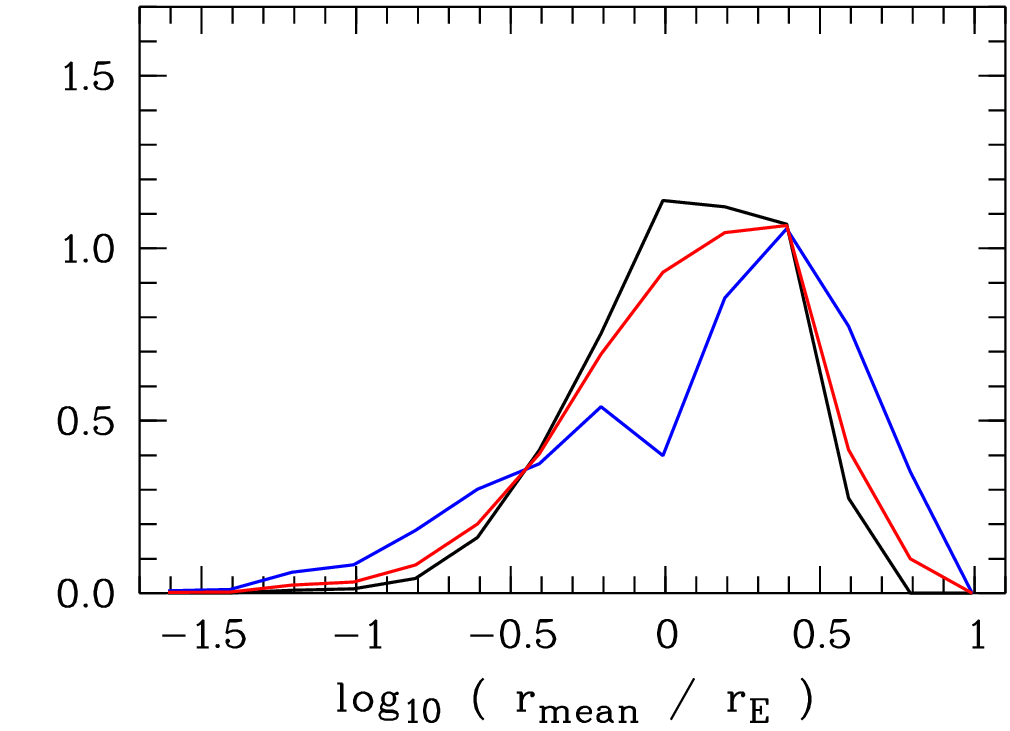}\includegraphics*{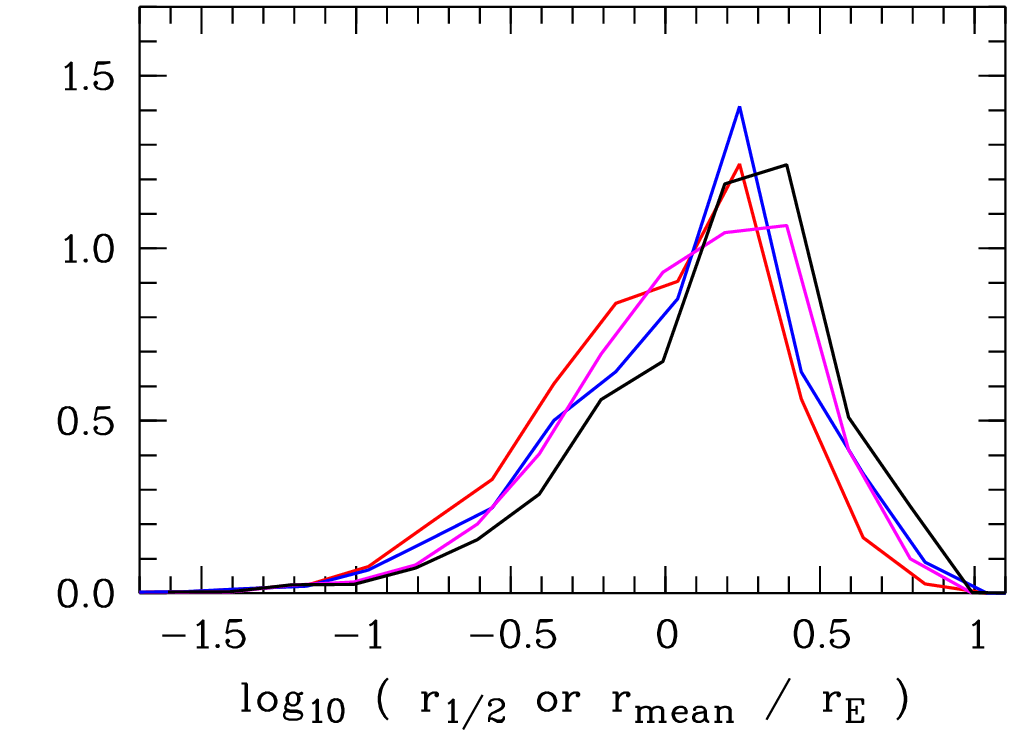}}
\caption{Posterior probability densities of the radius of the \ion{C}{iv} BLR in J1138. The BLR radius is expressed in Einstein radius units, with $r_E$ = 11.7 light-days for $\mathcal{M} =  0.3  \mathcal{M}_{\odot}$. The top panel probabilities are computed with $M(\text{\ion{C}{iv}})$ = 0.90, and the bottom panel probabilities with $M(\text{\ion{C}{iv}})$ = 1.05. {\it Left panels}: Probability densities of the half-light radius, $r_{\text{1/2}}$, obtained with two magnification maps that were computed with different fractions of compact objects, $\kappa_{\star} / \kappa $ = 7\% (black) and $\kappa_{\star} / \kappa $ = 20\% (blue), and after marginalizing over the two maps (red).  {\it Middle panels}: Same as the left panels but for the flux-weighted mean radius, $r_{\text{mean}}$. {\it Right panels}: Comparison of the probability densities, marginalized over the two maps, of the half-light radius (red and blue curves) and the flux-weighted mean radius (magenta and black curves), computed with constraints from the continuum source magnification (red and magenta curves) and without this constraint (blue and black curves).}
\label{fig:sizeblr2}
\end{figure*}

\section{Results}
\label{sec:results}

In Fig.~\ref{fig:fitmuv} we show examples of simulated $\mu(v)$ profiles that fit the observed $\mu(v)$ profiles. Although very different in J1339 and J1138, the observed profiles are clearly reproduced by a number of simulated profiles. The $\mu(v)$ profiles observed in J1004$+$4112 and Q2237$+$0305 were similarly reproduced \citep{2023Hutsemekers,2024Savic}, showing that, within the uncertainties, the diversity of $\mu(v)$ profiles can be reproduced with the simple BLR models under consideration, with no need for more complex geometries or kinematics.

\subsection{Probability of the different BLR models}
\label{sec:proba}

To identify the BLR models that best fit the observations, we computed the relative posterior probability that a given model $(G, i)$ (where $G$ = KD, PW, or EW, and $i$ = 22\degr, 34\degr, 44\degr, or 62\degr) can reproduce the  observables by marginalizing the likelihood over $r_s$, $r_{\text{in}}$ and $q$, as well as over the microlensing parameters. Since the different BLR models share the same parameters and associated priors, we quantified their relative efficiency to reproduce the data by comparing their likelihoods, that is, by normalizing the marginalized likelihood by the sum of the likelihoods associated with each model $G$ for each inclination $i$. This procedure yields the relative probability of the different models.

The results are given in Tables~\ref{tab:proba1} and~\ref{tab:proba2}. Since the preferred model can depend on the map orientation with respect to the BLR axis, we computed the probabilities for $\theta \leq 30\degr$ and $\theta \geq 60\degr$ separately. On the other hand, the preferred models are found to be essentially independent on the map parameters, in particular the fraction of compact matter, so that the probabilities computed for the different maps are merged. For J1339, the EW model is favored for all $\theta$, while PW has a low probability. For J1138, the preferred model, either KD or EW, strongly depends on the map orientation relative to the BLR axis, as in J1004$+$4112 \citep{2023Hutsemekers}. PW has a lower, although not negligible, probability.

With KD preferred in Q2237$+$0305\footnote{In Q2237$+$0305 the preferred \ion{C}{iv} BLR model does not depend on $\theta$.} \citep{2021Hutsemekers, 2024Savic}, EW preferred in J1339, and either KD or EW preferred in J1138 and J1004$+$4112 depending on the map orientation, we can conclude that the flattened (disk) geometries better reproduce the microlensing effects on line profiles, while the kinematics is dominated by either Keplerian rotation or equatorial outflow.   

\begin{table}[]
\caption{Probability of the \ion{C}{iv} BLR models in J1339.}
\label{tab:proba1}
\renewcommand{\arraystretch}{1.1}
\centering
\begin{tabular}{lccccccc}
\hline\hline
     & \multicolumn{3}{c}{$\theta \leq 30\degr$} &  & \multicolumn{3}{c}{$\theta \geq 60\degr$} \\
\hline
     \multicolumn{8}{c}{2014/05/20} \\
\hline
          & KD & PW & EW &  & KD & PW & EW  \\
\hline
          22\degr         & 13 &  0 &  4    & &  6 &  0 & 28  \\
          34\degr         &  4 &  0 & 16    & &  2 &  0 & 42  \\
          44\degr         &  3 &  0 & 16    & &  1 &  0 & 18  \\
          62\degr         &  6 &  5 & 33    & &  0 &  0 &  3  \\
          All $i$         & 26 &  5 & 69    & &  9 &  0 & 91    \\
\hline
     \multicolumn{8}{c}{2017/04/06} \\
\hline
          & KD & PW & EW &  & KD & PW & EW  \\
\hline
          22\degr         & 5 &  0 &  11    & &  3 &  0 &  44  \\
          34\degr         & 1 &  0 &  16    & &  1 &  0 &  30  \\
          44\degr         & 1 &  0 &  15    & &  0 &  0 &  17  \\
          62\degr         & 2 &  8 &  40    & &  0 &  0 &   5  \\
          All $i$         & 9 &  8 &  82    & &  4 &  0 &  96    \\
\hline
\end{tabular}
\tablefoot{The probabilities are given in percent, for the two epochs separately. $\theta$ is the angle between the BLR axis and the magnification map orientation.}
\end{table}

\begin{table}[]
\caption{Probability of the \ion{C}{iv} BLR models in J1138.}
\label{tab:proba2}
\renewcommand{\arraystretch}{1.1}
\centering
\begin{tabular}{lccccccc}
\hline\hline
     & \multicolumn{3}{c}{$\theta \leq 30\degr$} &  & \multicolumn{3}{c}{$\theta \geq 60\degr$} \\
\hline
     \multicolumn{8}{c}{$M(\text{\ion{C}{iv}})$ = 0.90} \\
\hline
          & KD & PW & EW &  & KD & PW & EW  \\
\hline
          22\degr         & 36 &  0 &  0    & &  0 &  6 & 16  \\
          34\degr         & 25 &  7 &  0    & &  0 &  8 & 25  \\
          44\degr         & 13 & 13 &  0    & &  0 &  9 & 16  \\
          62\degr         &  6 &  0 &  0    & &  0 &  6 & 13  \\
          All $i$         & 80 & 20 &  0    & &  0 & 29 & 70    \\
\hline
     \multicolumn{8}{c}{$M(\text{\ion{C}{iv}})$ = 1.05} \\
\hline
          & KD & PW & EW &  & KD & PW & EW  \\
\hline
          22\degr         & 32 &  0 &  0    & &  0 &  5 & 14  \\
          34\degr         & 21 &  7 &  0    & &  0 &  8 & 25  \\
          44\degr         & 11 & 23 &  0    & &  0 &  9 & 16  \\
          62\degr         &  5 &  0 &  0    & &  0 &  9 & 14  \\
          All $i$         & 69 & 30 &  0    & &  0 & 31 & 69   \\
\hline
\end{tabular}
\tablefoot{The probabilities are given in percent, for the two macro-magnification factors separately. $\theta$ is the angle between the BLR axis and the magnification map orientation.}
\end{table}

\subsection{Size of the broad emission line region}
\label{sec:sizeblr}

By marginalizing over all parameters but $r_{\text{in}}$, we can estimate the most likely BLR radius. Since $r_{\text{in}}$ does not properly represent the size of the BLR, which also depends on the light distribution, we computed the half-light and flux-weighted mean radii, $r_{\text{1/2}}$ and $r_{\text{mean}}$ respectively, for the different models, following \citet{2021Hutsemekers}. To compute the Einstein radius (Eq.~\ref{eq:re}), we adopted a flat lambda cold dark matter ($\Lambda$CDM) cosmology with $H_0 = 68$ km~s$^{-1}$ Mpc$^{-1}$ and $\Omega_m$ = 0.31. For J1339, the Einstein radius is $r_E$ = 10.1 $\sqrt{ \mathcal{M} / 0.3  \mathcal{M}_{\odot}}$ light-days. For J1138, $r_E$ = 11.7 $\sqrt{ \mathcal{M} / 0.3  \mathcal{M}_{\odot}}$ light-days.

Figures~\ref{fig:sizeblr1} and~\ref{fig:sizeblr2} show the posterior probability densities, uniformly resampled on a logarithmic scale, of the BLR radii $r_{\text{1/2}}$ and $r_{\text{mean}}$ for J1339 and J1138, respectively. The results obtained with the different magnification maps, computed with different values of the fraction of compact objects, $\kappa_{\star} / \kappa$, are illustrated separately. The median of the probability distributions is essentially independent of $\kappa_{\star} / \kappa$, while the distributions obtained with the highest $\kappa_{\star} / \kappa$ values are broader. A similar behavior was noticed for J1004$+$4112 \citep{2023Hutsemekers}. For J1339, the median of the probability density obtained with the high $\kappa_{\star} / \kappa$ value is shifted toward higher radii in 2017, but the distributions still largely overlap. 

Our microlensing simulations simultaneously fit the magnification profile $\mu(v)$ and the continuum magnification $\mu^{cont}$, considering a range of continuum source sizes (Sect.~\ref{sec:models}). For comparison, we also show, in the right panels of Figs.~\ref{fig:sizeblr1} and~\ref{fig:sizeblr2}, the probability distributions computed by only reproducing the $\mu(v)$ profile (which requires much less computing time). In most cases, the probability distributions are in excellent agreement, indicating that taking into account $\mu^{cont}$ does not improve the constraints on the BLR size. Also, neglecting the constraint from $\mu^{cont}$ does not change the relative probabilities of the BLR models given in Tables~\ref{tab:proba1} and~\ref{tab:proba2}.

Tables~\ref{tab:rad1} and~\ref{tab:rad2} give the BLR radii, computed from the median values of the probability distributions shown in Figs.~\ref{fig:sizeblr1} and~\ref{fig:sizeblr2}. The uncertainties correspond to the equal-tailed credible intervals that enclose a posterior probability of 68\%. The radii and intervals are then converted to a linear scale, multiplied by $r_E$, and expressed in light-days. For J1339, the two epochs of observation are considered separately, and no significant difference can be observed. For J1138, higher radii are obtained with the highest $M(\text{\ion{C}{iv}})$ value. This is expected since higher values of $M(\text{\ion{C}{iv}})$ lead to less magnified $\mu(v)$ profiles (Fig.~\ref{fig:muv2}), which naturally arise from larger BLRs. Although there is only a 15\% difference between the two values of $M$, the derived BLR radii differ by a factor of about 3. On the other hand, a 15\% decrease in the value of $M(\text{\ion{C}{iv}})$ in J1339, which corresponds to $M(\text{\ion{C}{iv}})$ = 0.29 when computed with  $M_{\text{B}} / M_{\text{A}} = 0.175$  \citep{2021Shalyapin} instead of  $M_{\text{B}} / M_{\text{A}} = 0.20$ (Sect.~\ref{sec:bels}), leads to a BLR half-light radius that is only 20\% smaller. To understand and roughly quantify the dependence of the BLR radius on the $M$ factor, we write $\mu_+ / \mu_- = M_- / M_+$ where $M_-$ and $M_+$ denote the low and high values of $M$, respectively,  and $\mu_-$ and $\mu_+$ represent the corresponding $\mu$ values.  Assuming $\mu = 1 + k \, (r_E / r)^{1/2}$ close to a caustic where $k$ is a constant \citep[e.g.,][]{1993Witt}, we derive $r_+/r_- = [(\mu_- -1)/(\mu_- \, M_- /M_+ \; -1)]^2$, which shows that, at low magnifications $\mu_-$, a rather small change of the macro-magnification factor $M$ can result in large BLR radius differences, as seen in the case of J1138. The sensitivity of the BLR radius to $M$ is much lower in J1339, thanks to the stronger microlensing effect.

\begin{table}[]
\caption{\ion{C}{iv} BLR radius in J1339.}
\label{tab:rad1}
\renewcommand{\arraystretch}{1.5}
\centering
\begin{tabular}{lcccc}
\hline\hline
 &  \multicolumn{2}{c}{2014/05/20} & \multicolumn{2}{c}{2017/04/06} \\
\hline
 &  $r_{\text{1/2}}$ & $r_{\text{mean}}$  & $r_{\text{1/2}}$ & $r_{\text{mean}}$  \\
\hline
  Map $\kappa_{\star} / \kappa $ = 11\%  & 4.8 $^{+3.4}_{-2.3}$  &  4.9 $^{+3.6}_{-2.4}$ & 5.4 $^{+4.2}_{-2.7}$ &  5.8 $^{+4.3}_{-2.8}$ \\
  Map $\kappa_{\star} / \kappa $ = 52\%  & 5.4 $^{+5.9}_{-3.6}$  &  5.6 $^{+6.0}_{-3.6}$ & 8.2 $^{+6.4}_{-5.1}$ &  8.4 $^{+6.9}_{-5.2}$ \\
  All maps                              & 5.1 $^{+4.6}_{-2.9}$  &  5.2 $^{+4.9}_{-2.9}$ & 6.7 $^{+6.0}_{-3.8}$ &  7.0 $^{+6.0}_{-3.9}$ \\
  All maps; $\mu(v)$ fit only           & 4.2 $^{+4.8}_{-2.5}$  &  4.6 $^{+5.3}_{-2.6}$ & 4.0 $^{+5.8}_{-2.6}$ &  4.3 $^{+5.6}_{-2.7}$ \\
\hline
\end{tabular}
\tablefoot{The BLR radii are given in light-days, for the two different epochs, and assuming an average microlens mass of 0.3  $\mathcal{M}_{\odot}$.}
\end{table}

\begin{table}[]
\caption{\ion{C}{iv} BLR radius in J1138.}
\label{tab:rad2}
\renewcommand{\arraystretch}{1.5}
\centering
\begin{tabular}{lcccc}
\hline\hline
 &  \multicolumn{2}{c}{$M(\text{\ion{C}{iv}})$ = 0.90} & \multicolumn{2}{c}{$M(\text{\ion{C}{iv}})$ = 1.05} \\
\hline
 &  $r_{\text{1/2}}$ & $r_{\text{mean}}$  & $r_{\text{1/2}}$ & $r_{\text{mean}}$  \\
\hline
  Map $\kappa_{\star} / \kappa $ = 7\%      &   5.2 $^{+4.1}_{-2.7}$ &  6.7 $^{+5.0}_{-3.2}$ & 12 $^{+11}_{-7}$  &  14 $^{+15}_{-8}$  \\
  Map $\kappa_{\star} / \kappa $ = 20\%     &   4.0 $^{+7.2}_{-2.4}$ &  4.9 $^{+8.3}_{-2.9}$ & 16 $^{+22}_{-13}$ &  19 $^{+24}_{-15}$ \\
  All maps                                &   4.9 $^{+4.9}_{-2.7}$ &  6.3 $^{+5.7}_{-3.4}$ & 12 $^{+13}_{-8}$  &  15 $^{+17}_{-9}$ \\
  All maps; $\mu(v)$ fit only             &   5.5 $^{+6.5}_{-3.3}$ &  6.9 $^{+7.9}_{-4.0}$ & 15 $^{+15}_{-10}$ &  19 $^{+19}_{-12}$ \\
\hline
\end{tabular}
\tablefoot{The BLR radii are given in light-days, for the two values of the macro-magnification factor, $M$, and assuming an average microlens mass of 0.3  $\mathcal{M}_{\odot}$.}
\end{table}

A good knowledge of the macro-magnification factor $M(\lambda)$ including the differential extinction, is thus needed to accurately measure the BLR radius from microlensing. For J1138, this ratio was measured using the MmD method \citep{2012Sluse}, which gives correct values of $M(\lambda)$ if both the continuum and the lines are either magnified or demagnified, and if at least a part of the emission line is not microlensed \citep{2010Hutsemekers}. Unfortunately, these assumptions, although reasonable, are difficult to verify a priori, and can be wrong in some cases (cf. J1004$+$4112 discussed in \citealt{2023Hutsemekers}). Similarly, the hypothesis that the BEL cores are not affected by microlensing \citep[e.g.,][]{2023Fian,2024Fian} is often verified but not always (cf. the case of Q2237$+$0305 in \citealt{2021Hutsemekers}). Ideally, the factor $M(\lambda)$ should be  measured from the flux ratio of spatially unresolved forbidden lines, which are essentially unaffected by microlensing since they originate in the extended narrow-line region, or from mid-IR to radio observations if the differential extinction can be estimated. 

Finally, since the BLR kinematics is characterized by an outflowing EW in J1339, and cannot be unambiguously attributed to Keplerian motion in J1138 (Sect.~\ref{sec:proba}), we did not derive the mass of the black hole in these quasars, as can be done on the basis of the measured BLR size, the line velocity width, and the virial theorem.

\subsection{Size of the continuum source}
\label{sec:sizecont}

\begin{figure}[t]
  \resizebox{\hsize}{!}{\includegraphics*{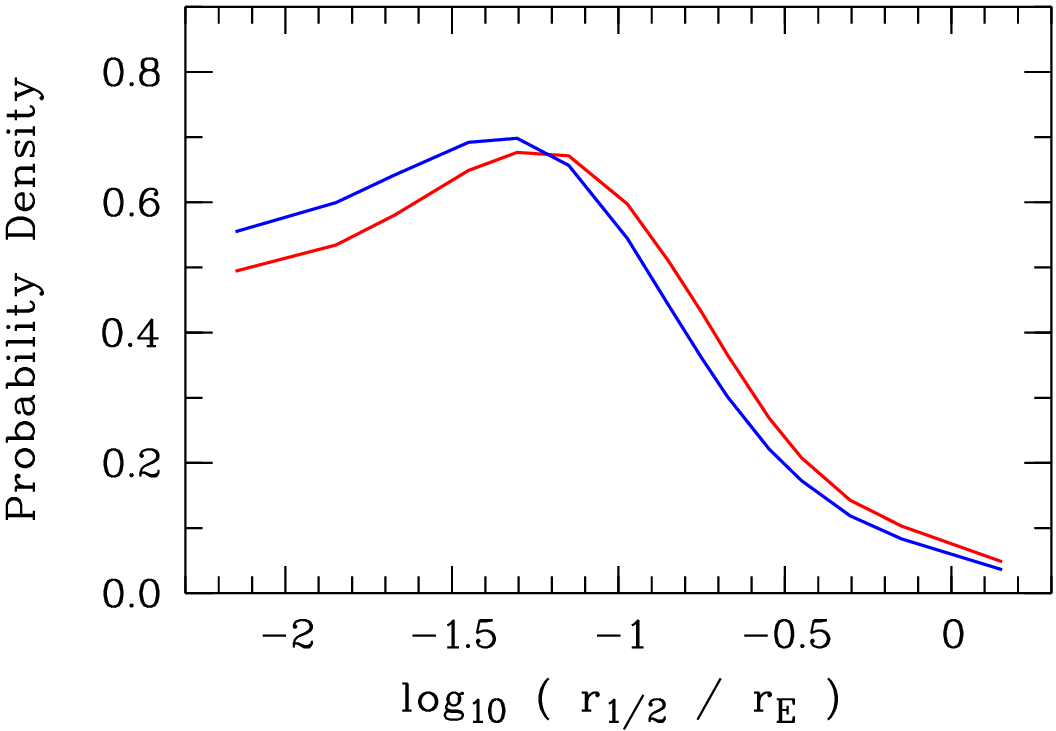}}
\caption{Probability distributions of the continuum source half-light radius in J1339, in units of $r_E$ = 10.1 light-days for $\mathcal{M} =  0.3 \mathcal{M}_{\odot}$, from the 2014 (red curve) and 2017 (blue curve) datasets.}
\label{fig:sizecont1}
\end{figure}

\begin{figure}[t]
  \resizebox{\hsize}{!}{\includegraphics*{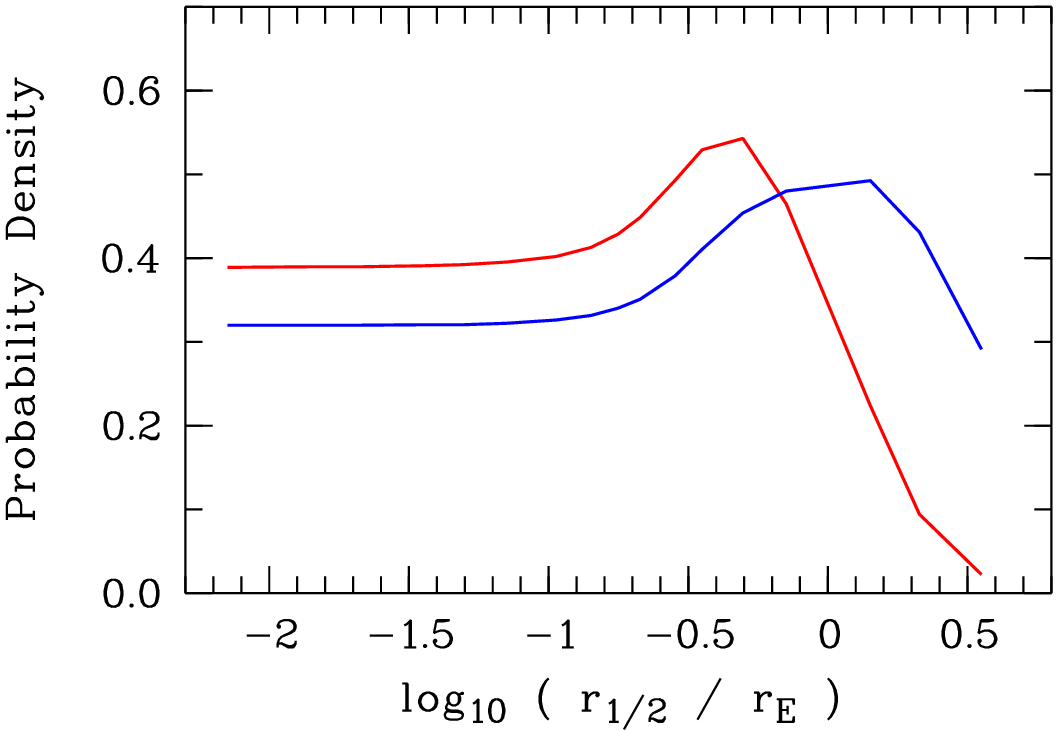}}
\caption{Probability distributions of the continuum source half-light radius in J1138, in units of $r_E$ = 11.7 light-days for $\mathcal{M} =  0.3 \mathcal{M}_{\odot}$, computed with $M(\text{\ion{C}{iv}})$ = 0.90 (red curve) and $M(\text{\ion{C}{iv}})$ = 1.05 (blue curve).}
\label{fig:sizecont2}
\end{figure}

By marginalizing over all parameters but $r_{\text{s}}$, including the different magnification maps, we estimated the most likely half-light radius of the continuum source, using  $r_{1/2} = r_s / \sqrt{2}$ for a uniform disk. The posterior probability distributions are shown in Figs.~\ref{fig:sizecont1} and~\ref{fig:sizecont2}. For J1339, there is a small shift toward smaller radii in 2017 that corresponds to the higher $\mu^{cont}$ value (Table~\ref{tab:indices}). A bigger shift is seen in the case of J1138 for the different values of $M(\text{\ion{C}{iv}})$, the highest $M(\text{\ion{C}{iv}})$ corresponding to a smaller $\mu^{cont}$ and, hence, a larger continuum size. Unfortunately, the distributions are not closed at small radii, so that only upper limits can be derived. This behavior, typical of single-epoch microlensing analyses, is due to the fact that similar magnifications can be obtained by changing either the source size or its position with respect to the caustics. This degeneracy can be removed with multi-epoch or multiwavelength data \citep[e.g.,][]{2011Blackburne}. We compute $r_{1/2} \lesssim 6$ light-days for J1339 and $r_{1/2} \lesssim 15$ light-days for J1138 ($M(\text{\ion{C}{iv}})$ = 0.9), with a posterior probability of 95\%. These upper limits are in agreement with  $r_{1/2}(\lambda_{\text{\ion{C}{iv}}}) \simeq 1$ light-day obtained for J1339 by \cite{2021Shalyapin} from $r$-band light curves, and $r_{1/2}(\lambda_{\text{\ion{C}{iv}}}) \simeq 4$ light-days obtained for J1138 by \cite{2011Blackburne} from chromatic microlensing.

\begin{figure*}[t]
  \resizebox{\hsize}{!}{\includegraphics*{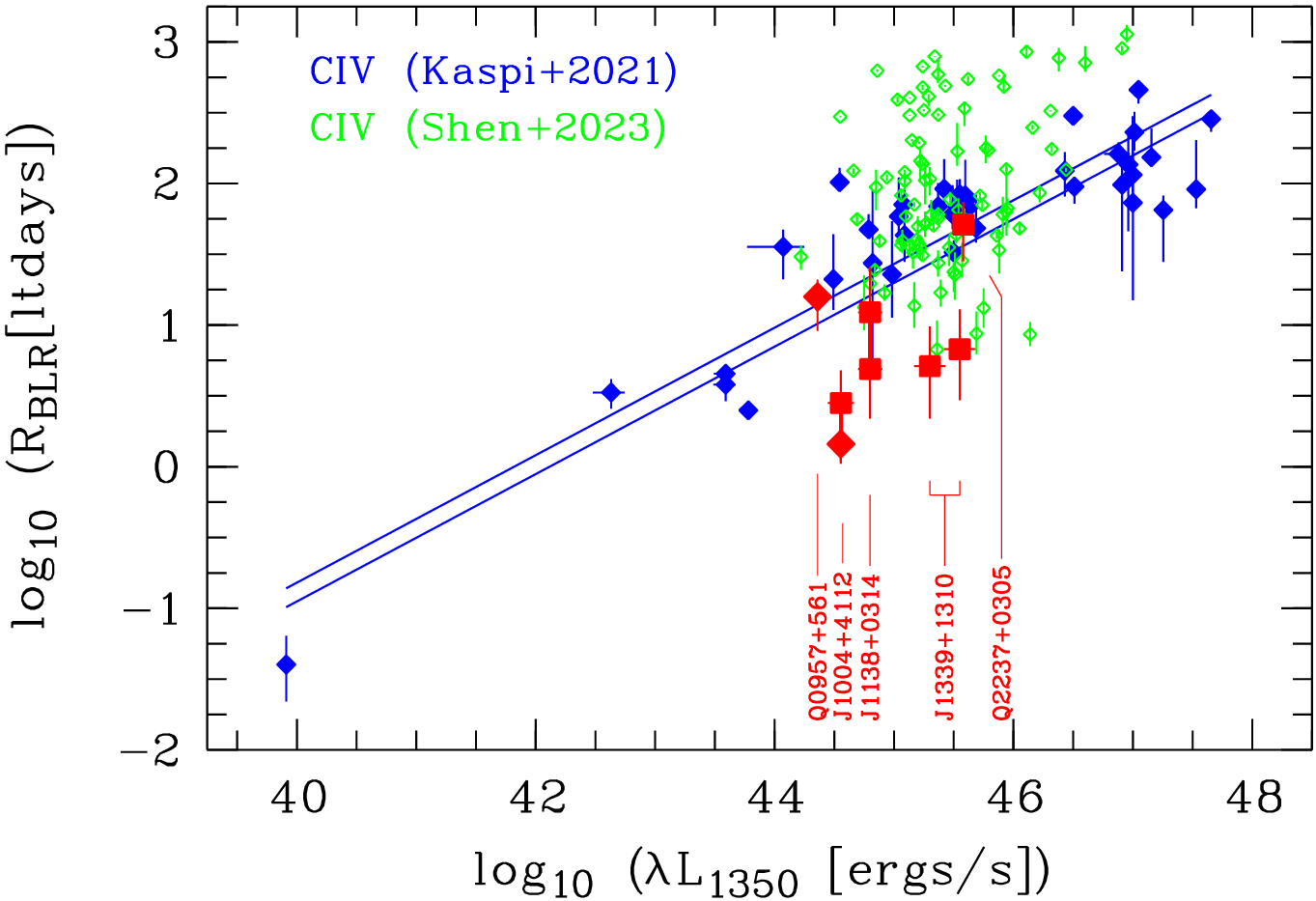}}
\caption{Radius--luminosity relation for the \ion{C}{iv} BLR. The rest-frame time lag from reverberation mapping and the continuum luminosity at 1350~\AA\ are from \citet[in blue]{2021Kaspi} and \citet[in green]{2023Shen}. Fits from \cite{2021Kaspi} are superimposed as continuous lines. The BLR half-light radii measured from microlensing are superimposed in red. Squares show the measurements from this work for J1339 and J1138,  and from \cite{2023Hutsemekers} and \cite{2024Savic} for J1004$+$4112 and Q2237$+$0305. Diamonds show the measurements from \cite{2023Fian,2024Fian} for Q0957$+$561 and J1004$+$4112.}
\label{fig:radlum}
\end{figure*}

\subsection{Radius--luminosity relation for the \ion{C}{iv} BLR}

\label{sec:radlum}

The $R$-$L$ relation for the \ion{C}{iv} BLR obtained from reverberation mapping is illustrated in Fig.~\ref{fig:radlum}, together with our measurements from microlensing. We used the half-light radii obtained by considering all magnification maps (Tables~\ref{tab:rad1} and~\ref{tab:rad2}), with the two epochs for J1339 and the two $M$ values for J1138 illustrated separately. \cite{2021Shalyapin} reported the luminosity of J1339 in 2014 and 2016, expressed in ergs~s$^{-1}$ units : $\log[\lambda L_{\lambda} (1350 \AA)]$ =  45.30$\pm$0.13 and 45.72$\pm$0.13, respectively. Since J1339 was fainter in 2017 than in 2016 (see Fig.~1 of \citealt{2021Shalyapin}), we estimated $\log[\lambda L_{\lambda} (1350 \AA)]$ =  45.55$\pm$0.15 in 2017. For J1138, we used $\log[\lambda L_{\lambda} (1350 \AA)]$ = 44.8$\pm$0.1 from \cite{2012Sluse}. The data for Q2237$+$0305 and J1004$+$4112 are taken from \cite{2021Hutsemekers}, \cite{2023Hutsemekers}, and \cite{2024Savic}. We emphasize that the BLR radius measured in Q2237$+$0305 has been estimated with both single-epoch data and times series, with values in excellent agreement \citep{2024Savic}.  Finally, we added the measurements independently obtained for Q0957$+$561 and J1004$+$4112 by \cite{2023Fian,2024Fian}, the value derived for J1004$+$4112 being in good agreement with ours.

The BLR radii obtained from microlensing seem to follow the global $R$-$L$ trend, although they are systematically on the lower side of the reverberation mapping $R$-$L$ relation; two quasars are exactly on the reverberation mapping $R$-$L$ relation, while two others are about one order of magnitude below. Depending on the value of $M(\text{\ion{C}{iv}})$, J1138 could be in either group. Interestingly, the BLR radius measured at two different epochs for J1339 apparently follows the trend, although the difference remains within the uncertainties. The fact that some values are significantly below the $R$-$L$ relation must be elucidated. A change of the average stellar mass from  $\mathcal{M} =  0.3 \mathcal{M}_{\odot}$ to $\mathcal{M} \gtrsim  3 \mathcal{M}_{\odot}$ could reconcile microlensing BLR radii with those from reverberation mapping, but such a high average stellar mass would be unrealistic \citep{2010Poindextera,2019Jimenez}. Since the intrinsic dispersion with respect to the $R$-$L$ relation is apparently very high, especially looking at the measurements of \cite{2023Shen}, a similar dispersion could be expected among the radii measured with microlensing, which could be biased toward small values by the selection of objects with particularly strong microlensing effects. On the other hand, this offset could be real and indicate that the BLR radius derived from microlensing or reverberation mapping does not exactly correspond to the luminosity-weighted radius \citep{2024Rosborough}. Finally, the offset could also indicate that a third parameter is required in the $R$-$L$ relation, likely the Eddington ratio as suggested for high-luminosity quasars \citep{2016Dub,2024Gravity}. More data are needed to investigate these issues in detail.

\section{Conclusions}
\label{sec:conclusions}

We analyzed the \ion{C}{iv} line profile distortions due to microlensing in two quasars, J1339 and J1138, for which high-quality spectra were available. This study complements the previous analyses of  microlensing-induced line deformations in the quasars Q2237$+$0305 and J1004$+$4112 \citep{2021Hutsemekers, 2023Hutsemekers, 2024Savic}.

J1339 shows a strong, asymmetric line profile deformation, while J1138 shows a more modest, symmetric deformation, confirming that microlensing can induce a rich diversity of spectral line distortions. To characterize the size, geometry, and kinematics of the \ion{C}{iv} BLR, we compared the observed line profile deformations to simulated ones. The simulations are based on three simple BLR models (KD, EW, and PW) of various sizes, inclinations, and emissivities, convolved with microlensing magnification maps specific to the quasar microlensed images. We conclude that:

\begin{itemize}
\item The various line profile deformations can be reproduced with the simple BLR models under consideration, with no need for more complex geometries or kinematics.
\item The models with disk geometries (KD and EW) are preferred, the PW being definitely less likely. In J1339, the EW outflow model is favored, while in J1138 the preferred model (KD or EW) depends of the orientation of the BLR axis with respect to the magnification map. In 
Q2237$+$0305, the KD rotating disk is more likely, indicating that different kinematics can dominate the \ion{C}{iv} BLR.
\item The measured half-light radius is $r_{1/2} =$ 5.1 $^{+4.6}_{-2.9}$  light-days for J1339 in 2014 and $r_{1/2} =$ 6.7 $^{+6.0}_{-3.8}$ light-days in 2017. They do agree within the uncertainties. For J1138, the amplitude of microlensing is smaller and more dependent on the macro-magnification factor. From spectra obtained in 2005 (single epoch),  $r_{1/2} =$ 4.9 $^{+4.9}_{-2.7}$  light-days or  $r_{1/2}= $ 12 $^{+13}_{-8}$ light-days, for two extreme, but reasonable, values of the macro-magnification factor.
\item The \ion{C}{iv} BLR radii from microlensing follow the radius--luminosity relation from reverberation mapping. However, while the sample of objects with microlensing measurements is yet small, there is possible evidence that microlensing radii lie, on average, below the $R$-$L$ relation. 
\end{itemize}

Further data are needed to confirm whether the BLR radii derived from microlensing are systematically smaller than radii obtained from reverberation mapping, and if this is due to a selection bias or a real difference. To secure accurate measurements from microlensing, a good estimate of the macro-magnification factor and differential extinction is required, especially when the microlensing magnification is small. We also find that different kinematics, rotation or outflow, can dominate the \ion{C}{iv} BLR. To understand how such a difference relates to other quasar properties, probing the BLR kinematics in a larger sample is definitely needed.

\begin{acknowledgements}
D.H. and Đ.S. acknowledge support from the Fonds de la Recherche Scientifique - FNRS (Belgium) under grants PDR~T.0116.21 and No 4.4503.19.
\end{acknowledgements}

\bibliographystyle{aa}
\bibliography{references}

\end{document}